\documentclass[iop]{emulateapj}
\usepackage{natbib}
\usepackage{amsmath}
\usepackage{hyperref}
\usepackage{subfigure}
\renewcommand\ion[2]{#1$\;${\scshape{#2}}}

\newcommand{\mcmc}[1]{#1}
\newcommand{\revision}[1]{#1}

\newcommand{\kms}{km s$^{-1}$}

\begin{document}

\title{Homologous Helical Jets: Observations by IRIS, SDO and Hinode \\ and Magnetic Modeling with Data-Driven Simulations}
\shorttitle{Homologous Helical Jets}

\author{Mark C. M. Cheung\altaffilmark{1},  B. De Pontieu\altaffilmark{1,2}, T. D. Tarbell\altaffilmark{1}, Y. Fu\altaffilmark{1,3}, H. Tian\altaffilmark{4}, \mcmc{P. Testa\altaffilmark{4},~} K. K. Reeves\altaffilmark{4}, \\J. Mart\'inez-Sykora\altaffilmark{1,5}, P. Boerner\altaffilmark{1}, J. P. W\"ulser\altaffilmark{1}, J. Lemen\altaffilmark{1}, A. M. Title\altaffilmark{1}, N. Hurlburt\altaffilmark{1},  L. Kleint\altaffilmark{6} , \\C. Kankelborg\altaffilmark{7}, S. Jaeggli\altaffilmark{7}, L. Golub\altaffilmark{4}, S. McKillop\altaffilmark{4}, S. Saar\altaffilmark{4}, M. Carlsson\altaffilmark{2} and V. Hansteen\altaffilmark{2}.}
\affil{1. Lockheed Martin Solar and Astrophysics Laboratory, 3251 Hanover Street Bldg. 252, Palo Alto, CA 94304, USA}
\affil{2. Institute of Theoretical Astrophysics, University of Oslo, P.O. Box 1029, Blindern, NO-0315 Oslo, Norway}
\affil{3. Space Sciences Laboratory, University of California, Berkeley, 7 Gauss Way, Berkeley, CA 94720, USA}
\affil{4. Harvard-Smithsonian Center for Astrophysics, 60 Garden Street, Cambridge, MA 02138, USA}
\affil{5. Bay Area Environmental Research Institute, 625 2nd St. Ste 209, Petaluma, CA 94952}
\affil{6. University of Applied Sciences and Arts Northwestern Switzerland, Bahnhofstr. 6, 5210 Windisch, Switzerland}
\affil{7. Department of Physics, Montana State University, Bozeman, P.O. Box 173840, Bozeman, MT 59717, USA}

\email{cheung@lmsal.com}

\begin{abstract}
We report on observations of recurrent jets by instruments onboard the Interface Region Imaging Spectrograph (IRIS), Solar Dynamics Observatory (SDO) and Hinode spacecrafts. Over a 4-hour period on July 21st 2013, recurrent coronal jets were observed to emanate from NOAA Active Region 11793. FUV spectra probing plasma at transition region temperatures show evidence of oppositely directed flows with components reaching Doppler velocities of $\pm 100$~\kms. Raster Doppler maps using a Si IV transition region line show all four jets to have helical motion of the same sense.  Simultaneous observations of the region by SDO and Hinode show that the jets emanate from a source region comprising a pore embedded in the interior of a supergranule. The parasitic pore has opposite polarity flux compared to the surrounding network field. This leads to a spine-fan magnetic topology in the coronal field that is amenable to jet formation. Time-dependent data-driven simulations are used to investigate the underlying drivers for the jets. These numerical experiments show that the emergence of current-carrying magnetic field in the vicinity of the pore supplies the magnetic twist needed for recurrent helical jet formation.
\end{abstract}


\maketitle

\section{Introduction}
 Since the discovery of coronal jets in X-ray~\citep[e.g.~][]{Shibata:1992Jet} and EUV~\citep[e.g.~][]{Chae:1999EUVJet} imaging observations, there has been a growing body of observational and theoretical work investigating the physical mechanisms behind this phenomena. While the detailed physical processes responsible for the acceleration of jet material depends on the local conditions (e.g. see Takasao et al. 2013), there is overwhelming evidence that magnetic reconnection is key for the impulsive energy release associated with jets. As for the driving mechanism that allows for energy build-up, it has been reported that many jets are associated with emerging flux and/or flux cancellation events in the photosphere. In the case of recurrent jets~\citep[e.g.][]{Chae:1999EUVJet,Chifor:2008Jet,Guo:RecurrentJets,Schmieder:TwistingJet,Chandra:SunspotWavesAndHomologousJets} emanating from the same source region on the Sun, an additional question is how the underlying driver leads to magnetic configurations that repeatedly erupt to produce jets of a homologous nature.

We address the question posed above by performing a study of recurrent jets observed by multiple spaceborne observatories and by using data-driven simulations. The rest of the article is structured as follows. Section~\ref{sec:obs_evidence} presents observations of the jets in the transition region and corona by the Atmospheric Imaging Assembly~\citep[AIA,][]{Lemen:AIA,Boerner:AIA} onboard the Solar Dynamics Observatory~\citep[SDO,][]{Pesnell:SDO} and by the Interface Region Imaging Spectrograph~\citep[IRIS,][]{DePontieu:IRIS}. Section~\ref{sec:physical_drivers}  presents photospheric observations in and around the source region of the jets by the Solar Optical Telescope~\citep[SOT,][]{Tsuneta:SOT} onboard the Hinode spacecraft~\citep{Kosugi:Hinode}.

Section~\ref{sec:hmi_vmag} describes the evolution of the photospheric field in the region of interest as revealed by vector magnetograms from the Helioseismic and Magnetic Imager~\citep[HMI,][]{Scherrer:HMI,Schou:HMI} onboard SDO. Section~\ref{sec:data_driven} presents results from simulations of coronal and chromospheric field evolution driven by HMI vector magnetograms. The physical implications of this study are discussed in section~\ref{sec:discussion}.

\section{Observational Evidence for Homologous, Helical Jets}
\label{sec:obs_evidence}
The recurrent jets and the source region of these jets were simultaneously observed by IRIS, SDO and Hinode. Instruments onboard these satellites provide complementary coverage in wavelength, temperature and the spatiotemporal domains. Together they present an integrated picture of the magnetic and atmospheric environment responsible for driving and initiating the set of recurrent jets. The following sections discuss observations of the region of interest as seen by the various instruments. 

\subsection{Photospheric and coronal observations by SDO}
\begin{figure*}
\centering
\resizebox{0.95\hsize}{!}{\includegraphics[width=0.9\textwidth]{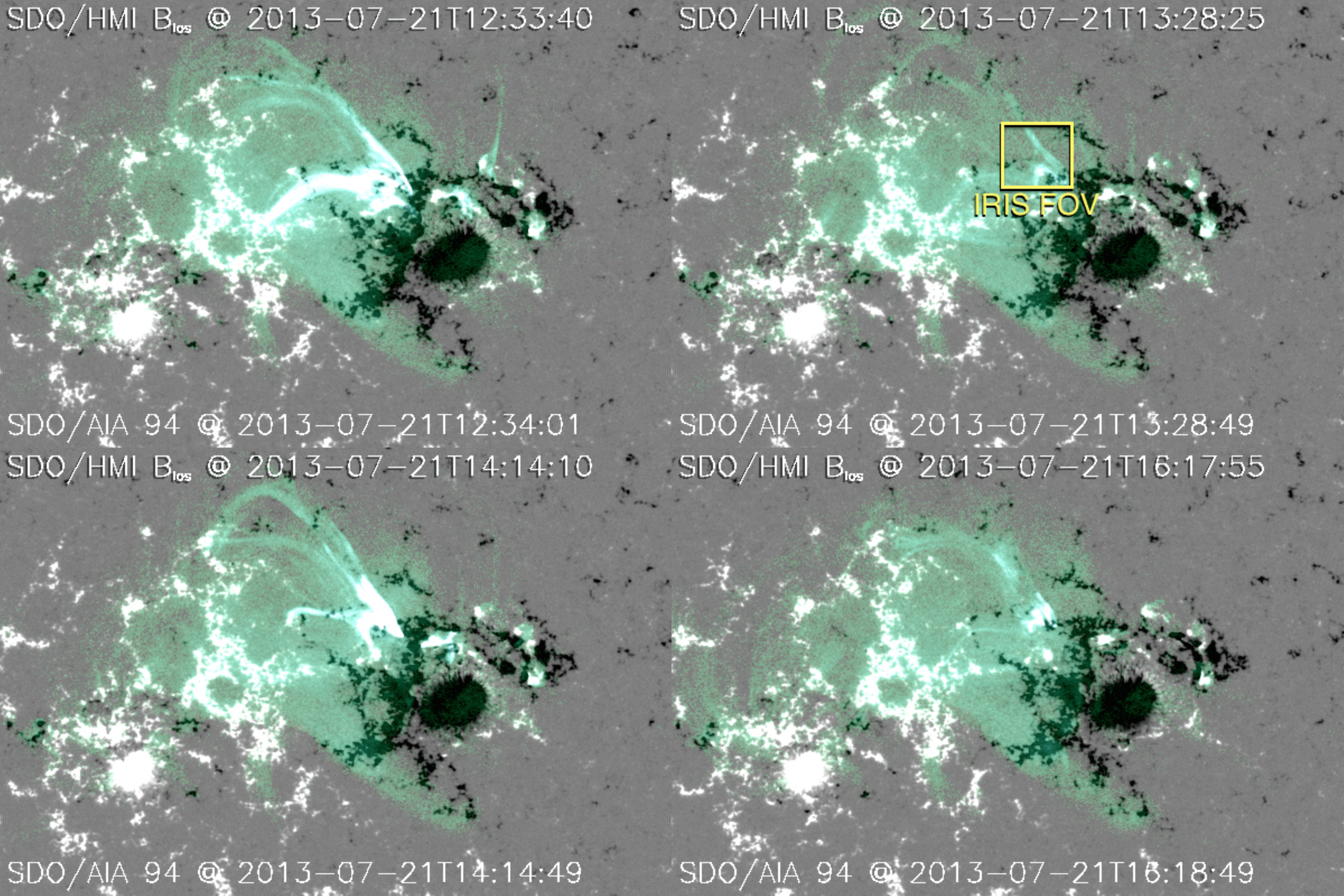}}
\caption{HMI line-of-sight magnetograms (greyscale) overlaid with AIA $94$~\AA~channel images of the four recurrent jets. The ejecta from the jets are channeled into closed loops connected the leading and trailing polarities of AR 11793. The yellow box in the top right panel shows the field of view of IRIS slit-jaw images.} \label{fig1}
\end{figure*}

Continuous full disk observations from SDO provide context about the environment in which the recurrent jets are generated. Fig.~\ref{fig1} shows the large-scale structure of active region (AR) 11793 in the time period during which IRIS observed the recurrent jets. During this time range the AR is roughly centered at a Stonyhurst longitude and latitude of W$11$ deg and N$23$ deg, respectively.

We refer to each of the four observed jets as J$1$, J$2$, J$3$ and J$4$ in chronological order. The four jets are shown in separate panels of Fig.~\ref{fig1}. Each panel consists of a HMI line-of-sight magnetogram (from the~\texttt{hmi.M\_45s} data series) overlaid with an EUV image from the AIA 94~\AA~channel.  The~\texttt{aia\_prep.pro} routine in SolarSoft was used to align the full disk images from the two instruments and to remap them to a common plate-scale of $0.6$ arcsec per pixel. 
Inspection of the overlaid images (and accompanying animation, available online) reveal that the jets emanate from a strong plage region in the northeastern edge of the leading (negative) polarity patch of the AR. 

Though the jets are not identical, they possess strikingly similar features. First of all, the ejecta in the jets are channeled into closed loops connecting the leading and trailing polarities of the AR. This is unlike some jets found in coronal hole regions, in which jet material is channeled into open magnetic field lines~\citep[e.g.][]{Cirtain:2007Jet,Savcheva:2007Jet,Moreno-Insertis:2008Jet,Patsourakos2008Jet}. Secondly, each of the jet events can be considered to possess a two-part structure, namely the inclined jet itself accompanied by the brightening of a compact closed loop (or multiple closed loops) adjacent to the jet~\citep{ShibataJet:1994,Shimojo:1996Jet}. It is worth noting that the jets can also be identified in the other EUV channels. This likely implies that they have multithermal structure.

In all four cases, the footpoints of the bright compact loops closest to the jets are found in the negative polarity network whereas the conjugate footpoints are located at a parasitic polarity patch within an adjacent supergranule. We will discuss the magnetic configuration of the environment around the jets in sections~\ref{sec:physical_drivers} and~\ref{sec:magnetic_field_evolution}.

\subsection{Transition region observations by IRIS}
\label{sec:tr_iris}
Between 11:34 and 16:34 UTC on 2013-07-21, IRIS ran a medium coarse 20-step raster observation program with $150$ repeats on the region of interest. The field-of-view of slit-jaw images (SJI) is $60''\times60''$ and the approximate pointing is indicated by the yellow box in the top right panel of Fig.~\ref{fig1}. The raster step size is $2''$ so each spectral raster spans a FOV of $38''\times 60''$. Each repeat of a spectral raster is accompanied by five slit-jaw images in each of the \ion{C}{ii} $1330$, \ion{Si}{iv} $1400$ and Mg II k $2796$  channels and one continuum image in $2832$. Level 2 data was used for all of the following analysis. Slit-jaw images in the level 2 data product are dark-subtracted, flatfielded and geometrically corrected so that images from different channels are on a common grid. The same corrections are applied to NUV and FUV spectra. Furthermore, the spectra are stacked as 3D raster cubes for convenient analysis.

\begin{figure*}
\centering
\includegraphics[width=0.9\textwidth]{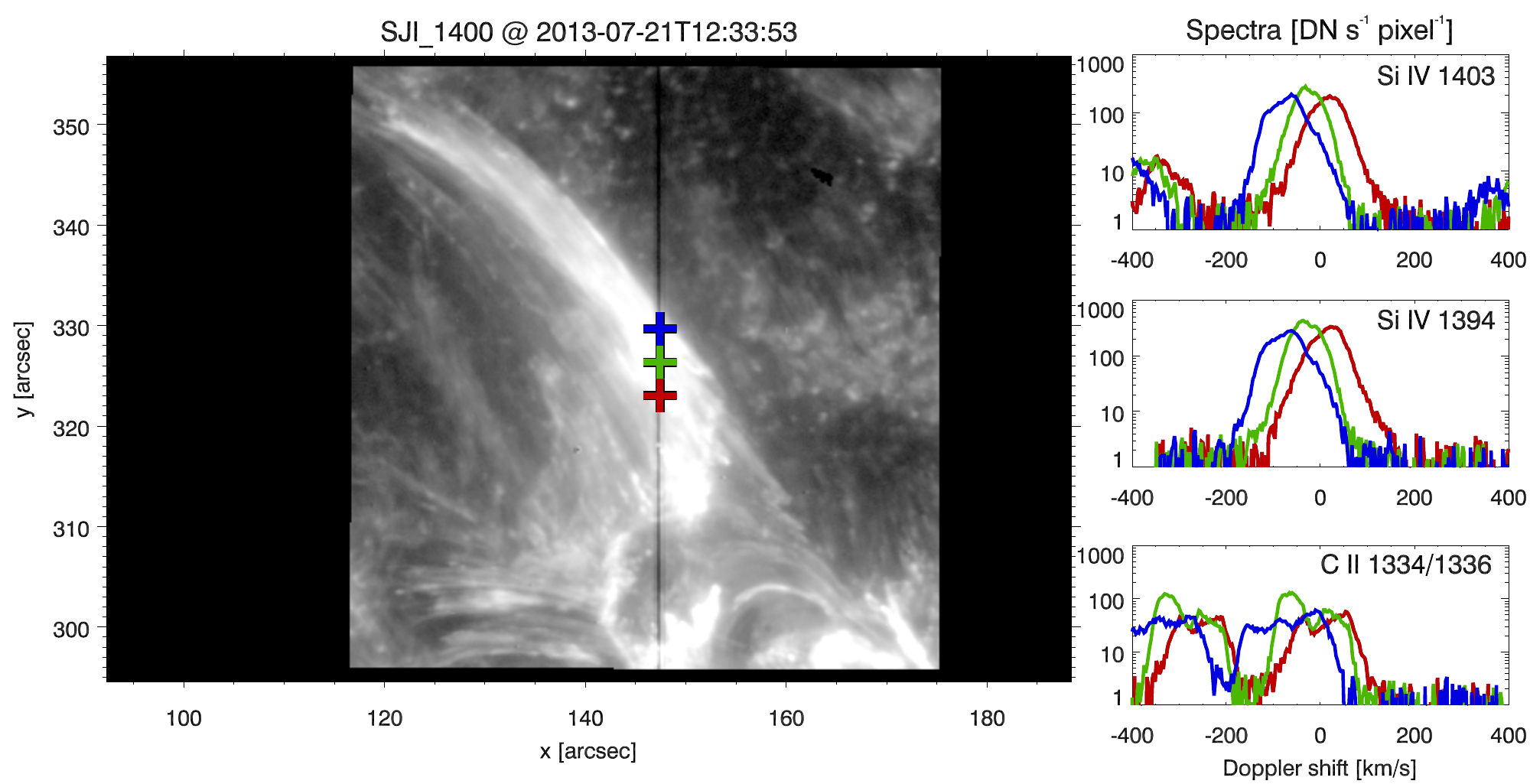}
\caption{Left: \ion{Si}{iv} 1400 slit-jaw image of the first recurrent jet. Right: Spectral line profiles sampled at the blue, green and red cursor positions (all lying on the vertical slit) for \ion{Si}{iv} 1403, \ion{Si}{iv} 1394 and the pair of \ion{C}{ii} lines at 1334 and 1336 (the rest wavelength of the latter is used as a reference). The spectra are plotted as functions of Doppler shift from their respective rest wavelengths.} \label{fig2}
\end{figure*}

Figure~\ref{fig2} shows a \ion{Si}{iv} 1400 slit-jaw image of J1. The spectrograph samples along the vertical dark slit at $x=154$ arcsec. FUV spectral line profiles are plotted for three different positions indicated by the blue, green and red cursors straddling the jet in the slit-jaw image. We first inspect the profiles of the \ion{Si}{iv} lines at $1393.8$ and $1402.8$ \AA. Both are transition region lines that form at $\log T/K \sim 4.9$~\citep[from CHIANTI 7.0,][]{CHIANTI7}. For this reason, the shapes of the line profiles are very similar. By visual inspection, it can be discerned that the \ion{Si}{iv} line profiles at the three positions have very different bulk Doppler shifts. While the centroids of the blue and green profiles have blueshifts of $\approx 80$ and $30$ km s$^{-1}$, respectively,  the centroid of the red profile is clearly redshifted. The spectral profiles at all three locations have wide wings with contributions beyond $\pm 100$ \kms~relative to the centroid positions. The spectral readout window of~\ion{Si}{iv} $1403$ spectra includes the \ion{O}{iv} $1401$ line, which forms at $\log T \sim 5.2$. The signal from this line appears in the plotted \ion{Si}{iv} $1403$ spectra at Doppler shifts beyond $-300$ \kms.

The bottom panel of the right column in Fig~\ref{fig2} shows spectra for the \ion{C}{ii} lines, which are expected to form at $\log T/K \sim 4.4$~\citep[from CHIANTI 7.0,][]{CHIANTI7}. In the plots the rest wavelength of the red line ($1335.71$~\AA) was used to calculate the effective Doppler shift. The shapes of the \ion{C}{ii} lines are much more complex than those of the \ion{Si}{iv} lines. The profiles at the green and red cursor positions are bimodal. However the green profile has a higher amplitude peak on the blue side of the line while the red line has a higher amplitude peak on the red side. Even more complex is the blue profile, which has a trimodal shape. One may initially be tempted to interpret the bimodal profiles in terms of oppositely directed Doppler flows. However, we caution that the \ion{C}{ii} lines are usually optically thick, so the dip in bimodal (or trimodal) line profiles likely results from opacity effects.

\begin{figure*}
\centering
\includegraphics[width=0.9\textwidth]{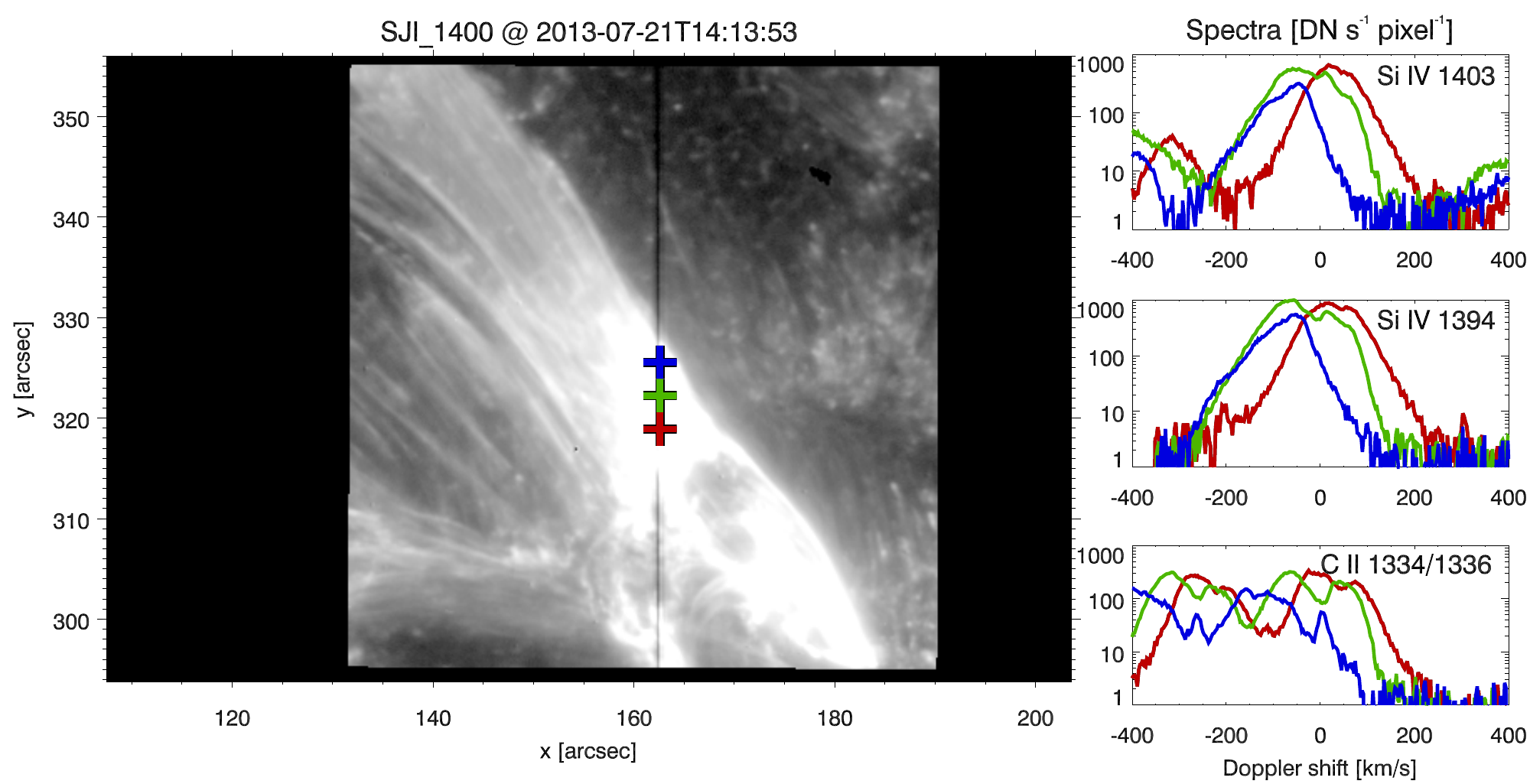}
\caption{Same as Fig.~\ref{fig2} but for the third jet. The FUV profiles in this case are even broader, with lines having contributions from up to $\pm 200$~\kms~Doppler shifts relative to the centroids.} \label{fig3}
\end{figure*}
\begin{figure*}
\centering
\includegraphics[width=0.9\textwidth]{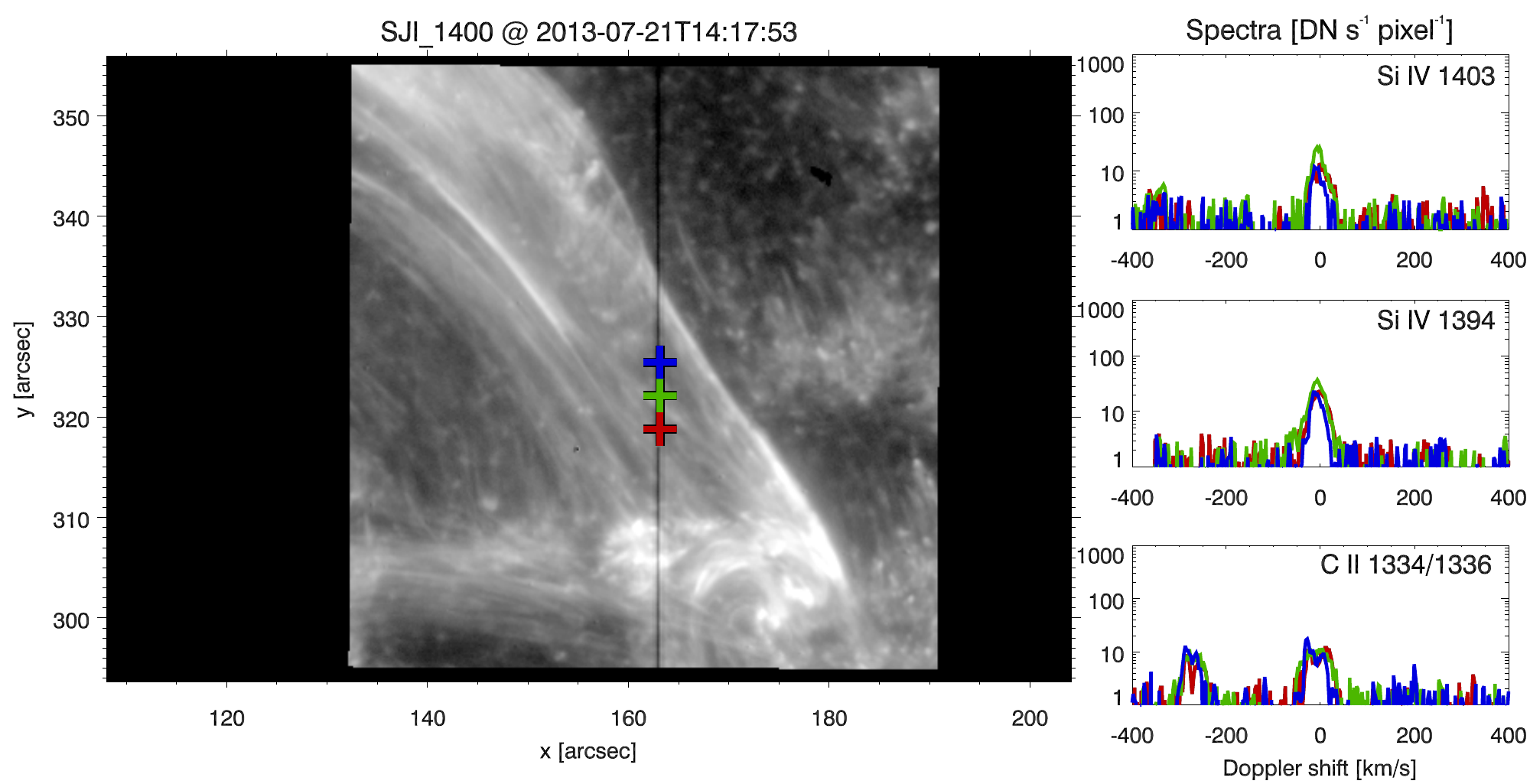}
\caption{Same as Fig.~\ref{fig3} but $4$ min later. The jet has already subsided and the FUV profiles are also much narrower than during the impulsive phase of the jet.} \label{fig4}
\end{figure*}

Fig.~\ref{fig3} is similar to Fig.~\ref{fig2} but for J3. There are many qualitative similarities between the profiles in these jets. For instance the red and blue colored profiles here show bulk red- and blueshifts, respectively. In the case of J3, the lines are even broader (with contributions up to $\pm 200$ \kms~relative to the centroid position) and the pair of \ion{C}{ii} lines are now completely blended. The green profiles for the \ion{Si}{IV} lines in Fig.~\ref{fig3} show a bimodal structure. This is probably different than the two-component spectral profiles studied by~\citet{Tian:CMEsCDsJets} in their study of EUV jets. In their case, they performed a double Gaussian fit to spectral profiles taken by the EUV Imaging Spectrograph (EIS) instrument onboard Hinode and found that one component corresponds to steady background emission, while a blue-shifted second component is attributed to outflows from the jet. In our case (green \ion{Si}{IV} 1394 profile in Fig. ~\ref{fig3}), one component has a bulk redshift while the other component has a bulk blueshift. Inspection of corresponding profiles at positions north and south of this location (i.e., as indicated by the blue and red cursors on the slit-jaw image) shows the profile to the north has a blueshifted component, while the profile to the south has a redshifted component. Since the profile sampled at the position of the green cursor is mid-way between the two, it is not surprising that the green spectral profile has both red- and blueshifted components.  In contrast, the FUV line profiles in the post-impulsive phase (four minutes later, see Fig.~\ref{fig4}) are dimmer by almost two orders of magnitude, are much narrower and have zero mean Doppler shift.

\begin{figure*}
\centering
\subfigure[Jet 1]{\includegraphics[width=0.49\textwidth]{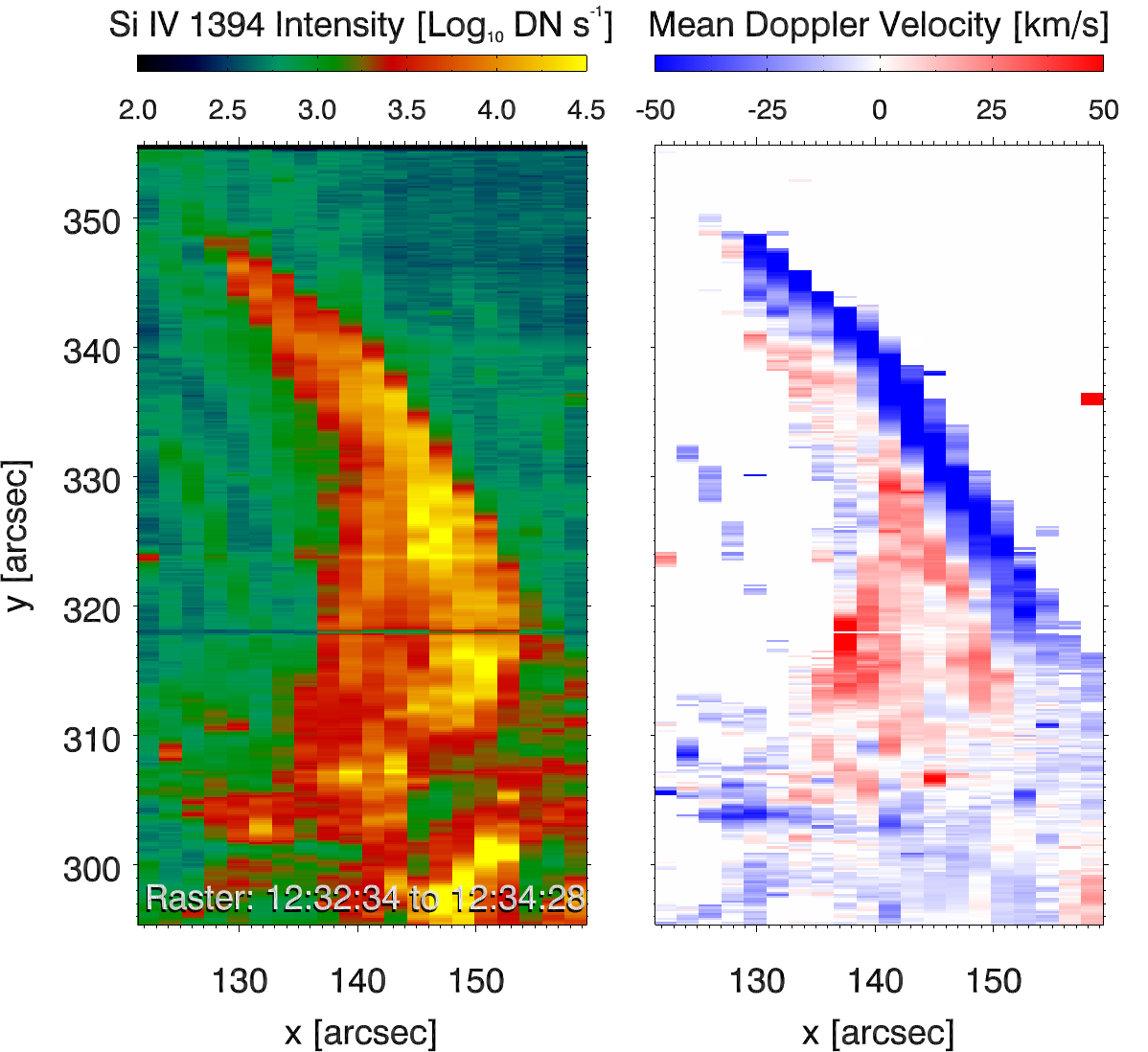}}
\subfigure[Jet 2]{\includegraphics[width=0.49\textwidth]{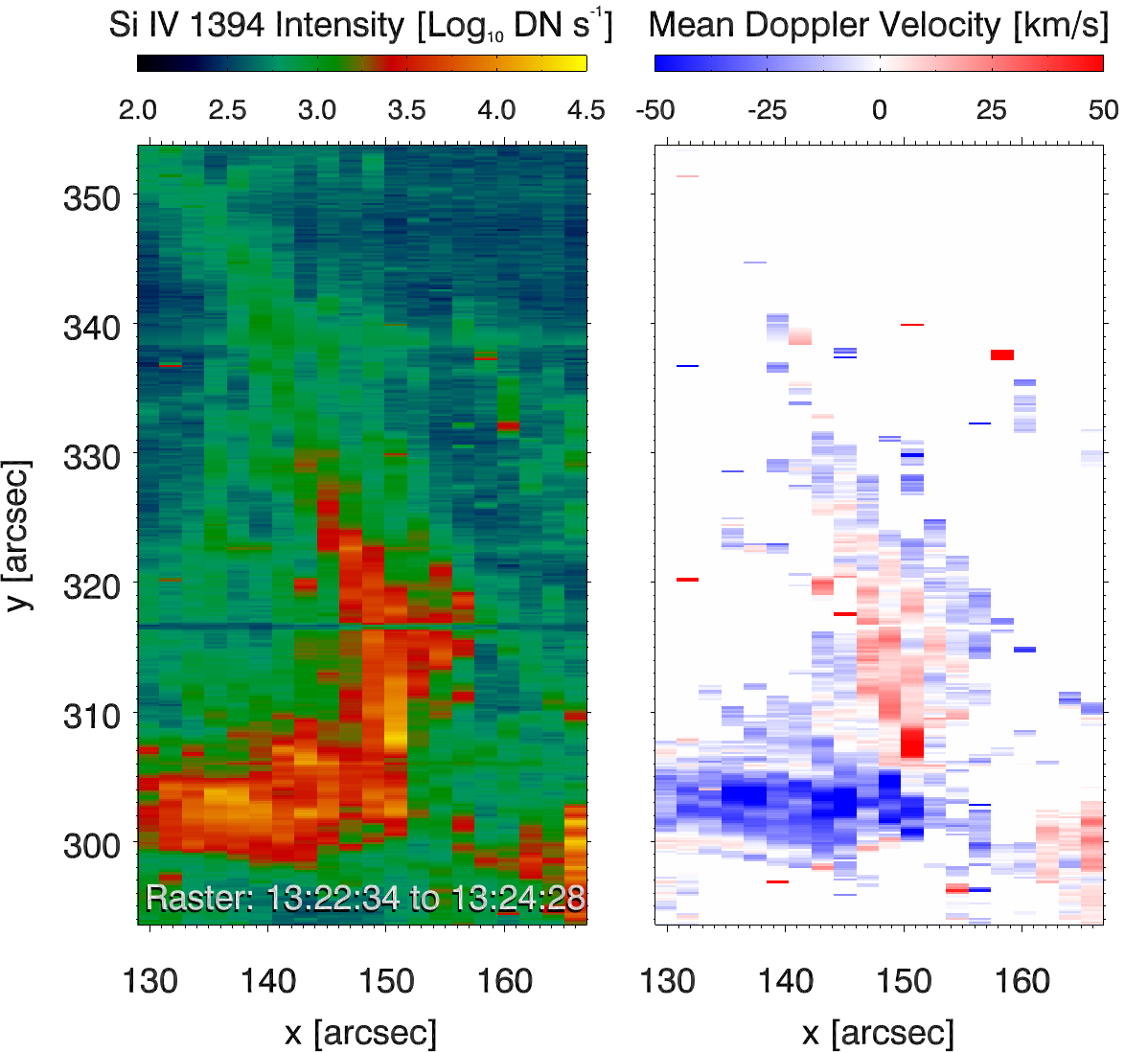}}\\
\subfigure[Jet 3]{\includegraphics[width=0.49\textwidth]{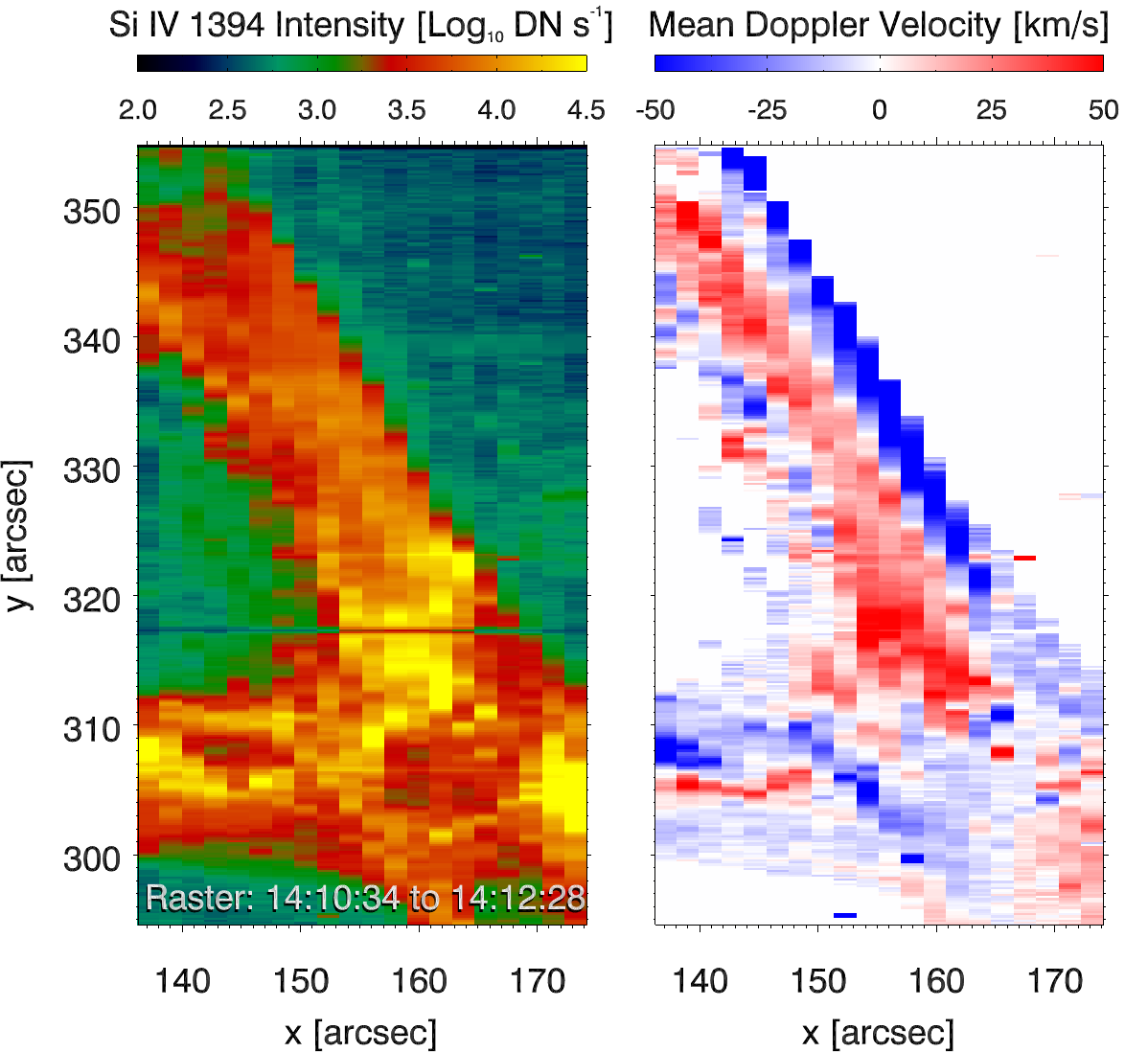}}
\subfigure[Jet 4]{\includegraphics[width=0.49\textwidth]{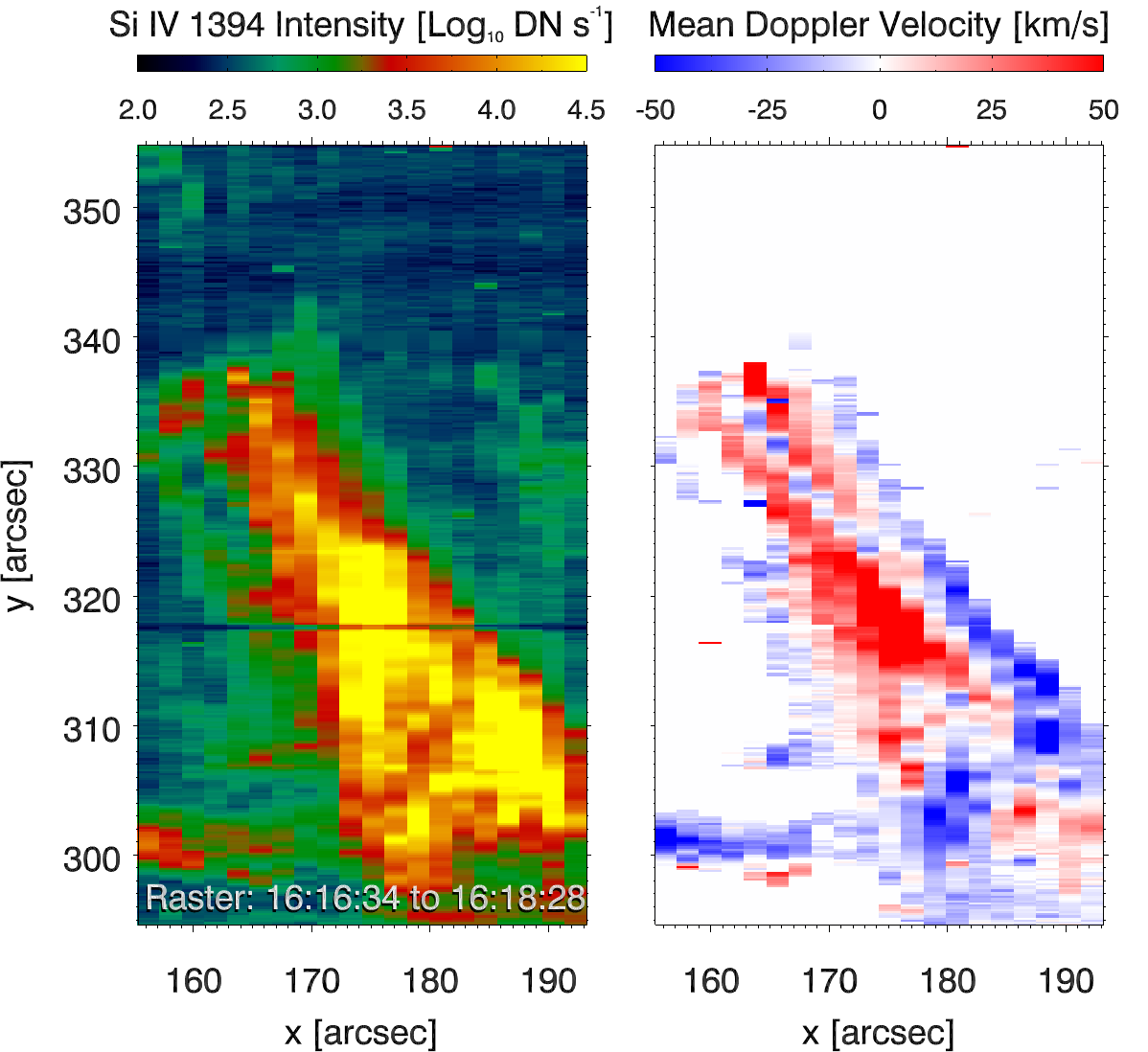}}
\caption{Total intensity and mean Doppler velocity of the four jets as computed from IRIS observations of the \ion{Si}{iv} $1394$ line. In all four jets, there is a tendency for northern edge to be blueshifted while the southern edge is redshifted. This spatial pattern suggests all four jets are helical with the same (negative) sign of kinetic helicity.}\label{fig5}
\end{figure*}

The FUV line profiles for different locations spanning the widths of J1 and J3 both indicate a bulk blueshift near the northern edge of the jet and a bulk redshift near its southern edge. Using raster scans, we investigate whether this is a pattern that pertains to all four jets. As per the above discussion, the \ion{Si}{iv} $1394$ line is simplest to analyze. Using this line, we calculate the zeroth and first moments
\begin{eqnarray}
I_{\rm line} &=& \int dI \\
\langle v_{\rm line} \rangle  &=& I_{\rm line}^{-1}\int I dv_{\rm los}, 
\end{eqnarray}
\noindent where $I$ is the spectrograph intensity (in DN sec$^{-1}$ pixel$^{-1}$) and $v_{\rm los}$ is the line-of-sight Doppler velocity. $I_{\rm line}$ is simply the integrated intensity across the line. Over the field-of-view of the IRIS rasters, the ratio $I_{1394} / I_{1403}$ is generally close to the theoretical value of $1.95$ as predicted using CHIANTI~\citep[][]{CHIANTI7}, which suggests the \ion{Si}{IV} lines are optically thin.  Adopting this assumption, we interpret $\langle v_{\rm line}\rangle$ to be an intensity-weighted mean Doppler velocity.

Figure~\ref{fig5} shows raster maps of $I_{\rm line}$ and $\langle v_{\rm line} \rangle $ for the \ion{Si}{iv} 1394 line for all four jets. While the amplitude of $\langle v_{\rm line} \rangle $ is larger in some jets than others, all four jets tend to have blue- and redshifts at their upper (northern) and lower edges (southern), respectively. This type of spatial pattern in Doppler maps have previously been detected in individual jets~\citep{PikeMason:RotatingJets} and surges~\citep{Curdt:HelicityInExplosiveEvents}. While~\citet{PikeMason:RotatingJets}  used data from the Coronal Diagnostic Spectrometer~\citep[CDS,][]{Harrison:CDS} and~\citet{Curdt:HelicityInExplosiveEvents} used data from the Solar Ultraviolet Measurements of Emitted Radiation~\citep[SUMER,][]{Wilhelm:SUMER} instrument, both studies used the same~\ion{O}{v} transition region line, which forms at $\log T/K \sim 5.4$. In both studies, the spatial pattern of Doppler shifts was taken as evidence for helical motion. We adopt the same interpretation and take our IRIS Doppler maps as evidence for helical motion in all four of the recurrent jets.  In a local Cartesian reference frame where $\hat{l}$ points along the jet (increasing height),  the spatial pattern of $\langle v_{\rm line} \rangle$ corresponds to rotational motion with vorticity $\omega_l =(\nabla\times\vec{v})_{l}< 0$.  This implies the kinetic helicity of plasma motion associated with the jet is $u_l \omega_l < 0$ (by definition, $u_l$ is positive since it is the component of plasma flow along the jet direction).

~\ion{O}{iv} lines observed in IRIS FUV spectra allow us to perform density diagnostics on the jet material. We used two different line ratios to measure the densities. The first is the ratio of the~\ion{O}{iv} $1401.1$~and $1404.8$~lines. The  $1404.8$~line is blended with a~\mcmc{\ion{S}{IV}} line, so we used the~\mcmc{\ion{S}{IV}~$1406.02$}~line and the assumption of optically thin emission to extract the intensity of~\ion{O}{iv} $1404.8$. The~\ion{O}{iv} $1404.8$~line is not always present in the spectral readout window of IRIS. However, in a number of slit positions where there is sufficient blueshift, we find the ratio \ion{O}{iv} $1401.1$/$1404.8$ to be in the range $4.0-4.5$. For a temperature range of $\log T/K = 4.5 - 5.5$ (derived from the ratio of~\mcmc{\ion{S}{IV}} 1404.8 to \mcmc{\ion{S}{IV}}~1406.02), this ratio gives densities ranging from $\log n_e/$ cm$^{-5} = 10.8 - 11.0$. Similarly, a ratio computed for the $1399$~and~$1401$~lines of~\ion{O}{iv} has values in the range $0.29-0.35$, which yields electron densities of $\log n_e/$ cm$^{-5} = 10.8- 11.2$.

\section{Physical Drivers of Recurrent Helical Jets}
\label{sec:physical_drivers}
Observations by SDO/AIA and IRIS establish the case that the jets are homologous and helical in nature. The four jets are homologous in the sense that they share substantial similarity in their observed spatial structure and evolution. This begs the question of the underlying driving mechanisms that lead to the initiation of the jets.

There is a large body of work using numerical MHD simulations to study how emergence of magnetic flux into pre-existing coronal field initiates jets~\citep{YokoyamaShibata:1995Jet,Miyagoshi:2004Jet,Archontis:EmergenceIntoCoronaII,Galsgaard:2005FluxEmergence,Isobe:EllermanBombs,Moreno-Insertis:2008Jet,Nishizuka:2008Jet,Heggland:2009Jet,Archontis:2010ReccurentJet,ArchontisHood:BlowoutJets,Moreno-Insertis:2013Jet,Takasao:2013Jet,Fang:RotatingSolarJets}. Most of the simulations focus on single jets following reconnection between the emerging magnetic system with the ambient field. Recent work started to investigate how multiple jets can be emitted from the same source region. From a 3D MHD simulation of flux emergence,~\citet{Archontis:2010ReccurentJet} reported that a series of reconnection events between the emerging flux system and ambient coronal field led to recurrent jets. \citet{Moreno-Insertis:2013Jet} performed a similar numerical experiment and found a succession of eruptions, some of which have physical properties that resembled the `standard' type of jet while others were miniature flux rope ejections that that may be associated with so-called blowout jets~\citep{Moore:2010JetDichotomy}. While the \citet{Moreno-Insertis:2013Jet} paper mentions that the erupting flux ropes in the simulation seem to rotate as if they were converting twist into writhe, the possible helical motion of the jets themselves were not studied.

\citet{Fang:RotatingSolarJets} modeled solar jets by performing 3D MHD simulations of twisted flux tubes emerging into a coronal layer with an ambient inclined field. They included magnetic field-aligned thermal conduction in their model, which provides the dominant mechanism for energy loss by plasma that has been heated to transition region and coronal temperatures. They reported the existence of columnar jets consisting of plasma at a broad range of temperatures (chromospheric to coronal). Due to acceleration by the Lorentz force acting on reconnected field lines, the jet columns exhibit spinning motion.  Synthetic intensity-weighted Dopplergrams (using coronal lines) of a simulated jet column from a side perspective gives net Doppler shifts of $\pm$ 20 km s$^{-1}$ on opposite sides of the columnar jet axis.

\mcmc{Recently,~\citet{Lee:HelicalBlowoutJets} performed a 3D MHD simulation of the emergence of a strongly twisted flux tube from the convection zone into the atmosphere. They imparted a density deficit distribution to the tube so that two segments of the tube would emerge and interact with each other as well as with the ambient inclined coronal field. The simulation yielded four episodes of twisted flux rope ejections carrying signicant mass loads away from the emerging flux region. These flux tubes untwist as they are ejected, resulting in signatures of \revision{torsional} Alfv\'en wave propagation. }

Apart from magnetic flux emergence, there is another means by which photospheric magnetic evolution can lead to jets, especially those with rotational motion.  One possibility of generating homologous helical jets was investigated by~\citeauthor{Pariat:2009Jet} (\citeyear{Pariat:2009Jet},~\citeyear{Pariat:2010Jet}). In their numerical MHD studies, the authors considered the scenario in which a circularly shaped patch of magnetic flux is embedded in an environment with predominantly vertical field of the opposite sign. For convenience let us call the circularly shaped patch of magnetic flux the `parasitic' pole. The initial potential field configuration has a coronal null with a fan that forms a quasi-separatrix (QSL). The QSL separates two magnetic volumes, one consisting of the set of closed field lines connecting the parasitic pole with its surroundings, and an exterior volume consisting of purely open magnetic field lines. About the axis of symmetry of this parasitic pole, they applied rotational motion at a fraction of the local Alfv\'en speed to mimic twisting of the field due to horizontal photospheric flows. The numerical simulations showed that, given sufficient cumulative rotation (greater than one turn), the magnetic system is driven to a state where reconnection of the magnetic field allows it to impulsively release the stored magnetic energy. A consequence of the relaxation is the formation of a helical jet. Subsequently, continued application of rotational driving at the bottom boundary led to the formation of similar helical jets.

The driving mechanism considered by Pariat et al. (2009, 2010) is in principle different to that of emerging flux in that only horizontal motions at the photosphere is applied. However, the emergence of magnetic flux can also drive systematic horizontal flows that shear already emerged field~\citep[][, section 3.6.3]{Manchester:ShearAlfvenWaves,Magara:HelicityInjection2003,Manchester:EruptionOfEmergingFluxRope,Magara:EmergingFluxSurfaceFlows,Cheung:ARFormation,Fang:FluxEmergence,Fang:DynamicCoupling,CheungIsobe:LivingReview}. Emergence-driven shear flows can lead to injection of magnetic energy and magnetic helicity for driving eruptive phenomena, including jets and CMEs. So the physical mechanism driving the recurrent jets reported here may not be exclusively due to one scenario or the other (i.e. flux emergence vs. shearing/rotational motion). 

To investigate the underlying physical driver(s) that cause the observed helical jets, we examine observations of the lower atmosphere by Hinode/SOT and SDO/HMI.

\subsection{Lower atmospheric observations by Hinode/SOT}
\label{sec:sot_obs}
\begin{figure*}
\centering
\includegraphics[width=1.\textwidth]{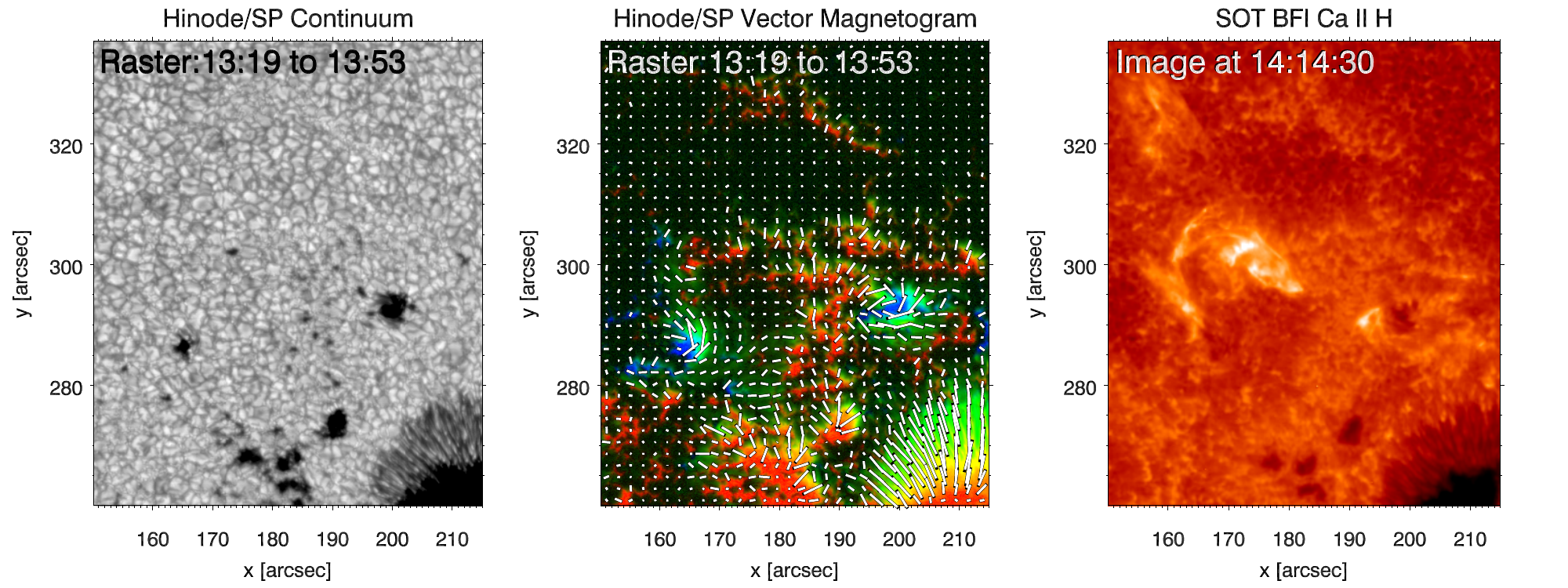}
\caption{Hinode SOT observations of the jet emitting region of AR 11793.~\emph{Left}: Continuum image from a Hinode/SP raster scan.~\emph{Center}: Vector magnetogram of the same field of view. The line-of-sight component ($B_{l}$) of the magnetic field is denoted such that blue and red denote positive and negative polarities, respectively. Green color coding shows the strength of the transverse component ($B_t$). Overplotted lines show the orientation of the transverse field. ~\emph{Right}: Broadband Filter (BFI) image in the Ca II H channel during the occurrence of the third jet.}
\label{fig6}
\end{figure*}
Figure~\ref{fig6} shows SOT observations of the jet emitting region of AR 11793. The left panel shows the continuum intensity image from a Hinode/SP raster scan. The central panel shows the vector magnetogram from a Milne-Eddington Stokes inversion~\citep{LitesIchimoto:HinodeSP,LitesEtAl:HinodeSP}. The right panel shows a broadband filter image from the~\ion{Ca}{ii} H channel during ocurrence of the third jet. The latter shows a set of closed loops and a set of inclined loops pointed toward the northeast direction. The morphology of the loops is somewhat reminiscent of the jet studied by~\citet[][see figure 1 of their article]{WeiLiu:UntwistingJet}. In their paper, they studied SOT observations of a jet using~\ion{Ca}{ii} H images alone. From the morphological evolution, they concluded that the apparent motion of the jet material was consistent with a helical jet with untwisting magnetic field lines. In the case they studied, the jet was observed off the solar limb so the two-part structure comprised of the closed loops and the inclined open loops were not contaminated by background emission. In our case, contribution to the SOT~\ion{Ca}{ii} H channel by emission in the upper photosphere makes it harder to delineate the specific features. Still, the morphological similarities with the case studied by~\citet{WeiLiu:UntwistingJet} support their conclusion that their jet exhibited helical motion.

The eastern (left) ends of the closed loops in the \ion{Ca}{ii} H image appear to be anchored at a compact patch of positive polarity field located at $(x,y)=(165'',287'')$. Inspection of the accompanying continuum image shows this positive patch to be a pore. This positive polarity pore is embedded in the interior of a supergranule. In contrast, the network field surrounding this supergranule is predominantly negative (the same polarity as the nearby leading spot). In this sense the pore is a parasitic pole. Inspection of the transverse field (i.e. the amplitude of the plane-of-sky component $|B_t|$) shows the presence of strong horizontal field ($B_t > 700$ G) on the western (right) side of the pore. Furthermore there is a moderate transverse field ($|B_t| \sim 300$ G) pervading a large fraction of the sugergranular cell. The left-right asymmetry of the $B_t$ distribution about the pore is suggestive of electric currents associated with a non-potential magnetic field configuration (more on this in section~\ref{sec:hmi_vmag}).

Inspection of the temporal sequence of \ion{Ca}{ii} H images indicates the presence of bright grain pairs that form in the vicinity of the parasitic pore. These bright grain pairs are separated by a dark lane with a length of $1-2$ Mm. This type of phenomena is a robust proxy for emerging flux ~\citep[e.g.][]{StrousZwaan:SmallScaleStructure,Cheung:SolarSurfaceEmergingFluxRegions,Guglielmino:FluxEmergence} and their presence in the vicinity of the parasitic pore suggests flux emergence may play a role in driving the recurrent jets.

\section{Magnetic Field Evolution}
\label{sec:magnetic_field_evolution}
In this section we investigate how the magnetic field in the source region of the jets evolve, and how this leads to recurrent jet production. For this purpose we use HMI vector magnetograms to drive simplified numerical models of chromospheric and coronal field evolution.

\subsection{HMI Vector Magnetograms}
\label{sec:hmi_vmag}
The Hinode/SP vector magnetogram shown in Fig.~\ref{fig6} suggests the presence of electric currents in the photospheric field around the parasitic pore. Due to Hinode telemetry limitations, only one SP map is available in the interval containing the four jets. So for the purpose of inspecting the evolution of the photospheric magnetic field, we used vector magnetograms from HMI instead.

\begin{figure*}
\centering
\subfigure{\includegraphics[width=\textwidth]{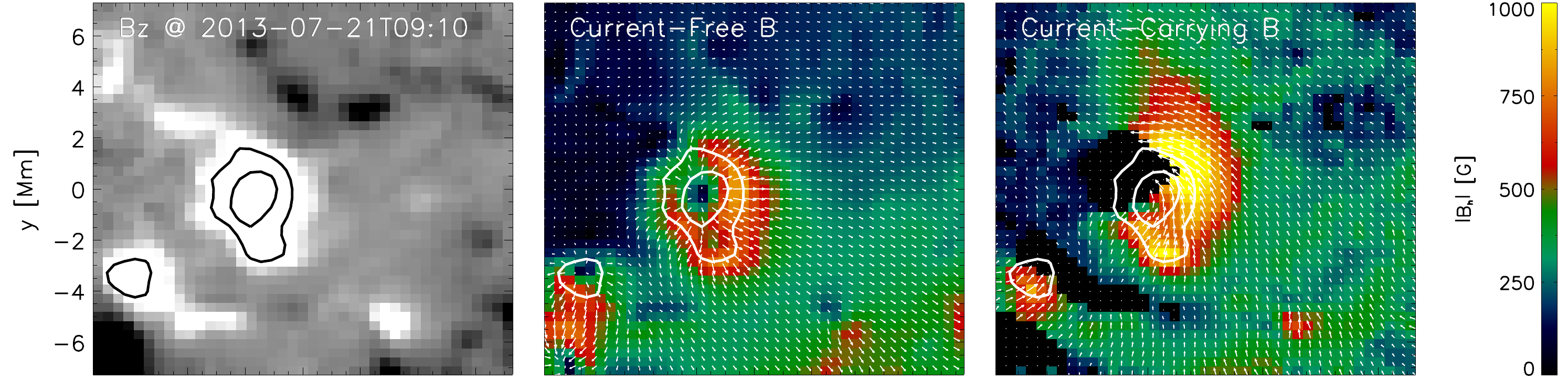}}\\
\subfigure{\includegraphics[width=\textwidth]{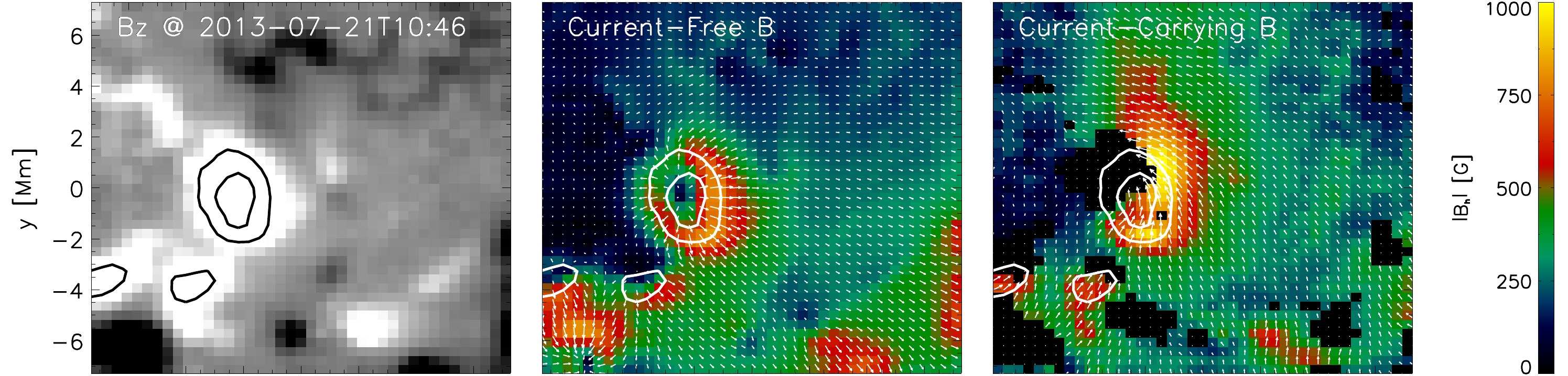}}\\
\subfigure{\includegraphics[width=\textwidth]{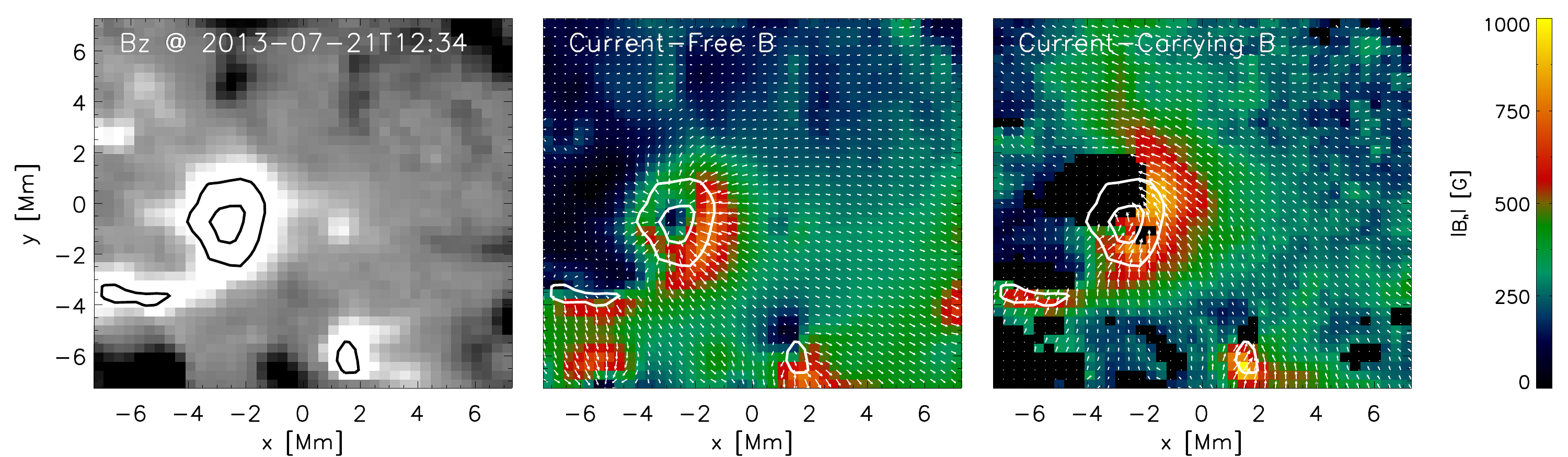}}
\caption{SDO/HMI vector magnetograms of the parasitic pore and its surroundings. \emph{Left}: Vertical component of $\vec{B}$ (greyscale saturated at $\pm 200$ Mx cm$^{-2}$). \emph{Middle}: Horizontal components of current-free (i.e. potential) part of $\vec{B}$. \emph{Right}: Horizontal components of the current-carrying part of $\vec{B}$. A strong, persistent patch of current-carry field is found on the west side of the pore. Comparison with the $B_z$ distribution shows this current-carrying patch is coincident with an emerging flux region (just northwest of the parasitic pore).  Contours for $B_z = 500$ and $1000$ Mx cm$^{-2}$ are shown on all panels to indicate the position of the pore. }\label{fig:vmags}
\end{figure*}

Each HMI vector magnetogram is produced by a Milne-Eddington inversion of IQUV Stokes maps temporally interpolated over an apodization window spanning $1350$ s~\citep{Hoeksema:HMI_Vmag}. This is done to increase the signal-to-noise ratio of the Stokes parameters and to filter out $p$-mode oscillations. While HMI vector magnetograms are not instantaneous representations of the photospheric field at any given time, the pixels in the same magnetogram are co-temporal. HMI provides vector magnetograms of the full AR at a regular cadence of one frame per $12$ min.  We use vector magnetograms from the HMI data series~\texttt{hmi.sharp\_cea\_720s}~\citep{Sun:SHARPs}. This series provides vector magnetograms in \revision{AR} patches, such that the magnetic field vectors have been disambiguated, transformed and remapped onto a cylindrical equal area grid. The magnetic vector is expressed as $(B_r,B_\theta,B_\phi)$, corresponding to the radial, longitudinal and latitudinal components, respectively. Since our main region of interest is relatively small ($L\sim 30$ Mm) , we ignore the effects of curvature and use the following mapping to a local Cartesian coordinate system $B_\theta \to B_x$, $B_\phi \to B_y$ and $B_r \to B_z$. This enables us to compute the vertical current density 
\begin{equation}
j_z = \frac{\partial B_y}{\partial x}-\frac{\partial B_x}{\partial y}.
\end{equation}
\noindent Inspection of maps of $j_z$ during the jet-emitting period reveals a persistent patch of positive $j_z$ near the parasitic (positive polarity) pore. Fig.~\ref{fig:vmags} shows HMI vector magnetograms in the neighborhood of the pore at 09:00, 10:46, and 12:34 UT. The left column shows the vertical component $B_z$. The positive polarity pore is roughly centered at $(x,y) = (-2,0)$ Mm in all three snaphots. The horizontal magnetic field can be decomposed into the sum of a potential component (calculated from $B_z$) and a current-carrying component. These are respectively shown in panels in the middle and right columns. \mcmc{A persistent patch of current-carrying field can be found on the west side of the pore at $(x,y)=(-1,0)$}. This feature is also found in the middle panel of Fig.~\ref{fig6}, which shows the Hinode/SP vector magnetogram of the same region. The Hinode/SP map provides a more accurate field measurement due to higher spectral coverage and resolution. The strength of the current-carrying horizontal field in this patch reaches $1$ kG and exceeds the field strength of the current-free counterpart. This indicates that the magnetic field on the west side of the pore is strongly twisted. Inspection of the $B_z$ time sequence from HMI shows that magnetic flux is emerging in this area. The orientation of the emerging flux is such that negative and positive polarity field are migrating in roughly the north and south directions, respectively. So both Hinode/SP and SDO/HMI vector magnetograms give evidence for the emergence of twisted field in the vicinity of the pore. In the following section, we present numerical simulations to investigate how this photospheric driving is related to the phenomena of recurrent helical jets. 

\subsection{Numerical Experiments using a Data-Driven Magnetofrictional Model}
\label{sec:data_driven}
We use a time-dependent magnetofrictional~\citep{YangSturrockAntiochos:Magnetofriction,CraigSneyd:Magnetofriction,vanBallegooijenPriestMackay:MeanFieldModel} model to carry out data-driven numerical experiments of coronal field evolution. Under this model, the fluid velocity $v$ appearing in the induction equation is assumed to be proportional to the local Lorentz force $\vec{j}\times\vec{B}$. This leads to an evolution of any arbitrary magnetic configuration to relax toward a force-free field. Magnetofriction has been used to model the formation and evolution of filaments~\citep{Mackay:FilamentChirality,vanBallegooijen:2004,MackayvanBallegooijen:2006a,Mackay:2009,Yeates:2007,Yeates:2008,Yeates:2009a}, the coronal field above the quiet Sun magnetic carpet~\citep{Meyer:2012} and the evolution of ARs~\citep{Cheung:DataDriven,Gibbs:AR10977}.

Following~\citet{Cheung:DataDriven}, we use a Cartesian magnetofriction code that solves for the vector potential $\vec{A}$, namely
\begin{equation}
\frac{\partial \vec{A}}{\partial t} = \vec{v}\times\vec{B},
\end{equation}
\noindent where $\vec{B}= \nabla \times \vec{A}$,~$\vec{j} = \nabla\times\vec{B}$ and 
\begin{equation}
\vec{v} = \frac{1}{\nu}\vec{j}\times\vec{B}.
\end{equation}
\noindent The magnetofrictional coefficient $\nu$ is given by
\begin{equation}
\nu = \nu_0 B^2 ( 1  - e^{-z/L}), 
\end{equation}
\noindent where $\nu_0 = 10$ s Mm$^{-2}$ and $L = 1.7$ Mm. As described in~\citet{Cheung:DataDriven}, the code uses a staggered grid (Yee mesh) such that $\vec{A}$ and $\vec{j}$ are defined at cell edges, $\vec{B}$ is defined on cell faces and $\vec{v} $ is defined at cell centers. The code has been updated to use a van Leer slope limiter~\citep{vanLeer:SlopeLimiter} to interpolate $\vec{v}$ onto cell edges when computing the $-\vec{v}\times\vec{B}$ electric field. We find that this scheme provides better numerical stability while being less diffusive than explicitly imposing an anomalous resistivity.

\subsubsection{Initial and boundary conditions}

\mcmc{The observed recurrent jets occur in the neighborhood of ambient inclined field as part of a set of AR loops that connect the leading and following polarities of AR 11793. To capture the large-scale magnetic connectivity, we use a computational domain sufficiently large to encompass the entire AR. The domain spans $248$ and $131$ Mm in the $x$ (longitudinal) and $y$ (latitudinal) directions, respectively. The bottom boundary is located at $z=0$ and the top boundary is located at $105$ Mm. The horizontal and vertical grid spacings are $364$ and $547$ km, respectively. The initial condition is a potential field of the AR computed for the magnetogram at 2013-07-21T06:12 UT (5 h 22 m before the start of the IRIS observation) and the simulations are evolved forward in time from that state.}

As discussed in~\citet{Cheung:DataDriven}, the bottom boundary condition for the magnetofrictional model is given by the transverse components of the photospheric electric field, namely $-\partial_t A_x = E_x$ and $-\partial_t A_y = E_y$. The retrieval of the horizontal electric field from vector magnetograms is a difficult inverse problem~\citep{Fisher:EstimatingElectricFields,Fisher:EfieldsWithDoppler}.  As recently demonstrated by~\citet{Kazachenko:2014}, the incorporation of Doppler flows as a constraint leads to inversion results that accurately reproduce electric fields (and the associated Poynting flux) in an anelastic MHD simulation. The validation of this method for use with vector magnetograms at the resolution and sensitivity of HMI on photospheric magnetic structures is work in progress.

Instead of attempting to perform a faithful retrieval of the photospheric electric field for this problem, we use a different method to compute electric fields with a given assumption.  Given the sequence of input vertical magnetograms $B_z$, its relation to $\vec{E}_h = (E_x,E_y)$ is given by the vertical component of the induction equation
\begin{equation}
\frac{\partial {B}_z(x,y) }{\partial t} = -\hat{z}\cdot ( \nabla \times \vec{E}_h).
\label{eqn:curlE}
\end{equation}
\noindent In order to solve for $\vec{E}_h$, another relation must be specified. Specifying the horizontal divergence of the electric field:
\begin{equation}
D(x,y) = \nabla_h \cdot \vec{E}_h.\label{eqn:divE}
\end{equation}
\noindent fully constrains the problem. In our numerical experiments, we assume 

\begin{equation}
D(x,y) = j_z(x,y)  U_0,\label{eqn:adhoc}
\end{equation}
\noindent \mcmc{where $j_z = (\nabla \times \vec{B})_z$ is the vertical current density computed at the photosphere. This choice of the functional form for $D$ is motivated by the following scenario. Consider an axisymmetric twisted magnetic flux tube that is invariant along its axis and let the tube axis be parallel to the vertical direction $\hat{z}$. Let the tube rise vertically with $\vec{v} = U_0 \hat{z}$. It can be shown that the divergence of the $-\vec{v}\times\vec{B}$ electric field corresponding to this motion is given by Eq.~(\ref{eqn:adhoc}). By adopting Eq.~(\ref{eqn:adhoc}), we have chosen to inject twist via the bodily emergence of twisted field~\citep[c.f.][]{Leka:CurrentCarryingFlux}. For driving the numerical experiments described here, $j_z$ is computed from HMI vector magnetograms and $U_0$ is a free parameter (with dimensions of velocity). Varying $U_0$ changes the effective injection speed of magnetic twist (as described by $j_z$) into the computational domain.}

\revision{Another possible way to inject twist is by means of shearing and rotational motions in the photospheric plane}. In MHD models of the emergence of twisted flux tubes, such flows are accelerated by magnetic torques exerted by the Lorentz force~\citep[][]{LongcopeWelsch:ModelForEmergence,Manchester:EruptionOfEmergingFluxRope,Magara:EmergingFluxSurfaceFlows,Fan:EmergenceOfATwistedFluxTube,Cheung:ARFormation,Fang:DynamicCoupling,CheungIsobe:LivingReview}. Twist injection by both bodily emergence of current-carrying field and rotational motions in the photosphere are likely present in our region of interest~(see section~\ref{sec:sot_obs}). For simplicity, however, we assume in our numerical setup that twist injection is due to the former.\footnote{For the specific case of an axisymmetric flux tube with azimuthal field $B_{\theta}(r) = q r B_l(r)$, where $q$ is the twist parameter and $B_l(r)$ is the longitudinal component of the magnetic field, one can pick parameters for the two scenarios (i.e.~bodily transport of twisted field and rotational motion) that result in identical boundary conditions.  For such a tube rotating about its axis with an angular velocity $\omega_0$, it can be shown that $\nabla \cdot \vec{E}_h = -q^{-1}\omega_0 j_z$, which is equivalent to Eq. (\ref{eqn:adhoc}) with $U_0 = -q \omega_0$.}

\subsubsection{Results of Numerical Experiments}
\begin{figure*}
\centering
\includegraphics[width=\textwidth]{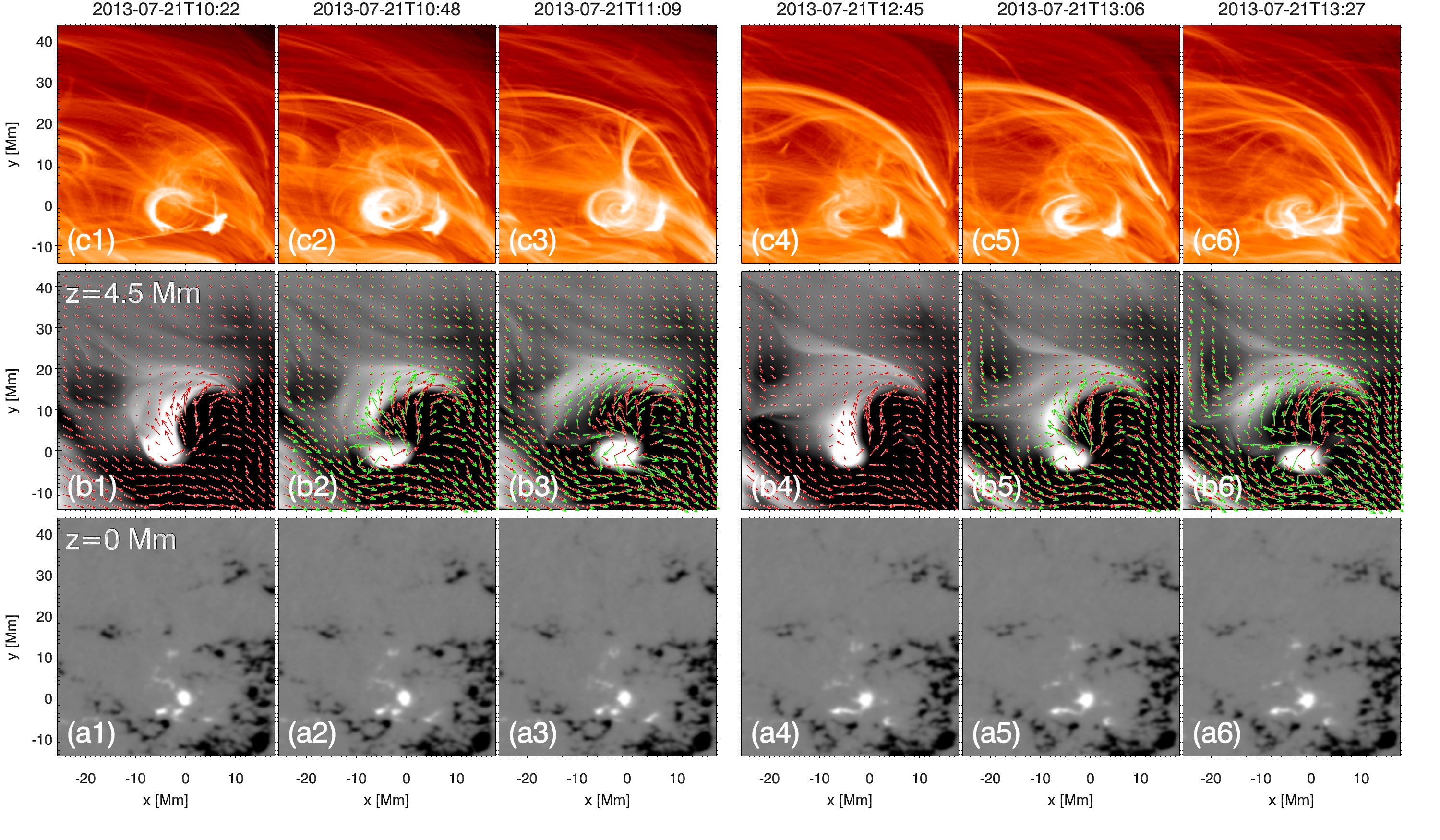}
\caption{Two examples of untwisting magnetic field in the data-driven model with $U_0 = 1.1$ km s$^{-1}$. Columns (1) - (3) show a temporal sequence of one such episode while columns (4)-(6) show a homologous episode later in time. In this local coordinate system the parasitic pore is centered at the origin. Row (a) shows the photospheric vertical field $B_z(z=0)$ scaled between $\pm 800$ Mx cm$^{-2}$. Row (b) shows synthetic chromospheric vector magnetograms at a constant height of $z=4.5$ Mm. Greyscale shows $B_z$ scaled between $\pm 100$ Mx cm$^{-2}$ and red arrows show the horizontal field at the same time. In the left and right sets of panels, the green arrows indicate the horizontal field at 10:48 and 12:45 UT, respectively (i.e. they show the horizontal field at the beginning of each set). The clockwise rotation of the arrows in both cases shows the magnetic field evolved with a clockwise rotation. Row (c) shows a visualization of the field lines according to the proxy emissivity model of Eq. (\ref{eq:emissivity}). }\label{fig7}
\end{figure*}

We carried out three numerical experiments, one each for $U_0 = 0$, $U_0 = 1.1$ and $U_0 = 2.2 $ km s$^{-1}$. For the run with $U_0 = 0$, $\vec{E}_h$ is decoupled from the photospheric distribution of $j_z$. In this case there is no systematic current injection and we do not find any magnetic field evolution resembling the helical motion of the observed jets. For $U_0 = 1.1$ km s$^{-1}$, we find repeated episodes of twisting followed by untwisting in the model magnetic field that shares qualitative similarities with the recurrent jets. Fig.~\ref{fig7} shows snapshots of two such homologous episodes of evolution from the model. Row (a) shows $B_z$ at the bottom boundary of the domain, which is constrained to match the $B_z$ observed by HMI. Row (b) shows synthetic chromospheric ``magnetograms'' (i.e. $B_z$ sampled at $z=4.5$ Mm). We chose to sample the magnetic field at a height that is above the nominal height range associated with 1D models of the solar chromosphere~\citep[$\sim 1$ to $2$ Mm, e.g. see][]{WithbroeNoyes:ChromosphereAndCorona,VALIII} since 3D radiative MHD simulations show that magnetic flux emerging into the atmosphere can lift the chromosphere and transition region layers by a few Mm~\citep{MartinezSykora:TwistedFluxEmergence}.  In both cases, the earliest chromospheric magnetogram (i.e. panels (b1) and (b4) ) shows a positive polarity feature pressed against a negative polarity feature in a ying-yang pattern. Inspection of the horizontal vectors and the distribution of $j_z$ at that height shows that the field near the polarity inversion line is sheared and carries current. When we inspect subsequent vector magnetograms, we find the horizontal vectors have, on average, rotated in a clockwise direction. This is a result of the untwisting of the magnetic field by a clockwise rotational flow. The vertical component of vorticity for such a flow is negative, consistent with the sign of vorticity of the observed recurrent jets. 

To visualize the magnetic field lines, we calculate a scalar proxy emissivity $\varepsilon(x,y,z)$ using the following procedure. For each field line traced from a position at the photospheric ($z=0$) boundary, we compute the following field-line averaged quantity:
\begin{equation}
\langle {\rm WD} \rangle  = L^{-1}\int_0^L \vec{F}_{\rm Lorentz} \cdot \vec{v} dl,
\end{equation}
\noindent where $L$ is the length of a field line and $\vec{F}_{\rm Lorentz} \cdot \vec{v} $ represents the rate of work done by the Lorentz force. In a full MHD model, $\vec{F}_{\rm Lorentz} \cdot \vec{v} $ appears as a source term in the kinetic energy equation and as a sink term (with negative sign) in the magnetic energy equation. In a magnetofrictional model, 
\begin{equation}
\vec{F}_{\rm Lorentz} \cdot \vec{v} = \frac{1}{4\pi\nu} [(\nabla\times\vec{B})\times\vec{B}]^2 \ge 0. 
\end{equation}
\noindent so that $\langle {\rm WD}\rangle \ge 0$.  If a field line crosses one of the side or top boundaries of the computational domain, we set $\langle  {\rm WD} \rangle=0$ so that the field line is not emissive. A magnetic field line will cross a number of cell elements in the computational domain. For each of these cell elements, we increment the local value of the emissivity by
\begin{equation}
d\mathcal{\varepsilon} = G \langle  {\rm WD}  \rangle\Delta x \Delta y,\label{eq:emissivity}
\end{equation} 
\noindent  where $G$ is some arbitrary scale-factor (here we use $G=1$). Row (c) shows integrals of the resulting proxy emissivity along vertical lines-of-sight. This method allows us to inspect a large number of fields lines (here we traced 4 field lines per pixel), each lit up according to the field-line averaged rate of magnetic energy loss due to work done by the Lorentz force. The images shown in row (c) are log-scaled so they do not depend on the specific value of the constant $G$. 

\begin{figure}
\centering
\subfigure[$B_z$ at $z=0$ Mm]{\includegraphics[width=0.4\textwidth]{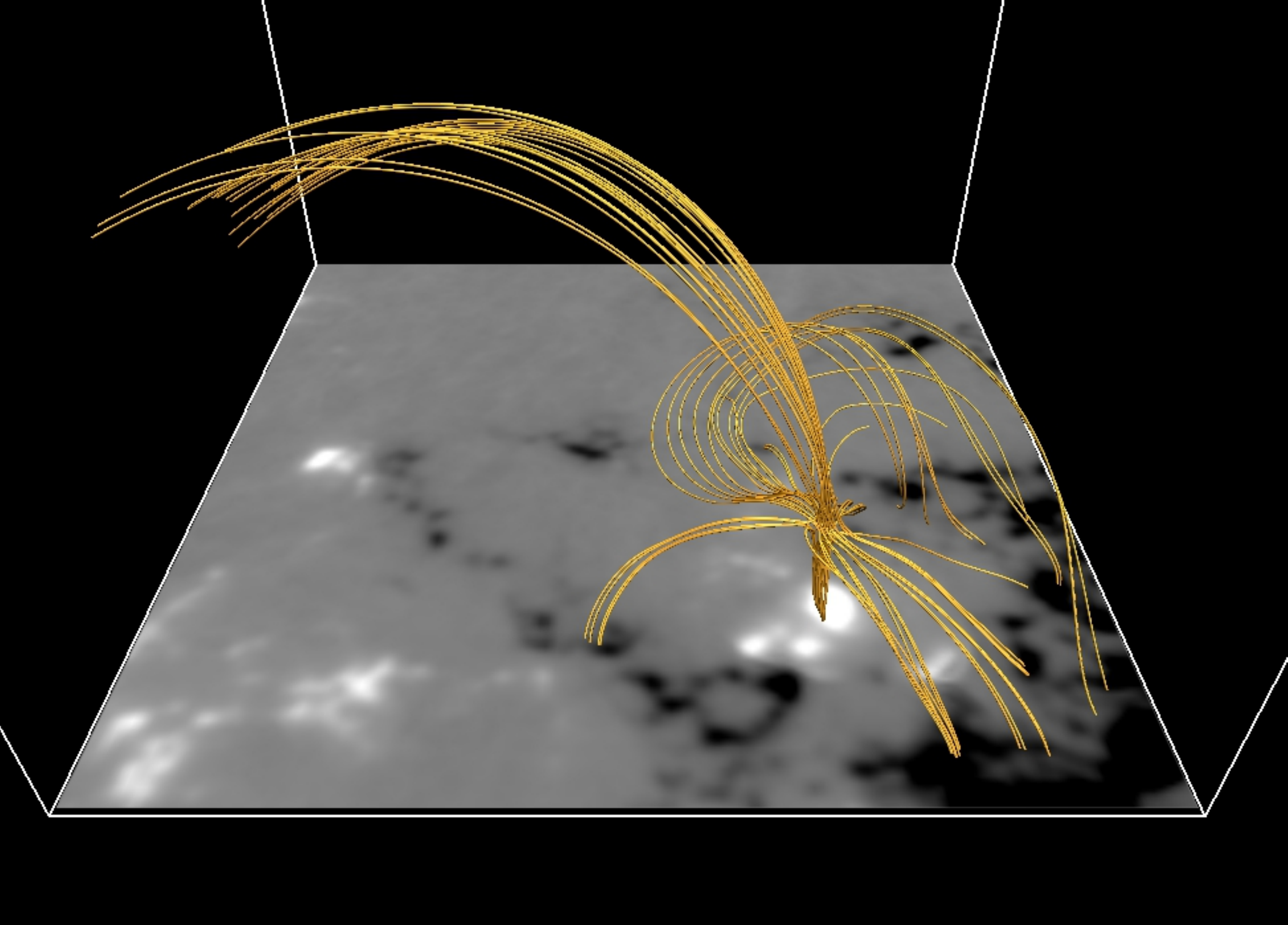}}\\
\subfigure[$B_z$ at $z=4$ Mm]{\includegraphics[width=0.4\textwidth]{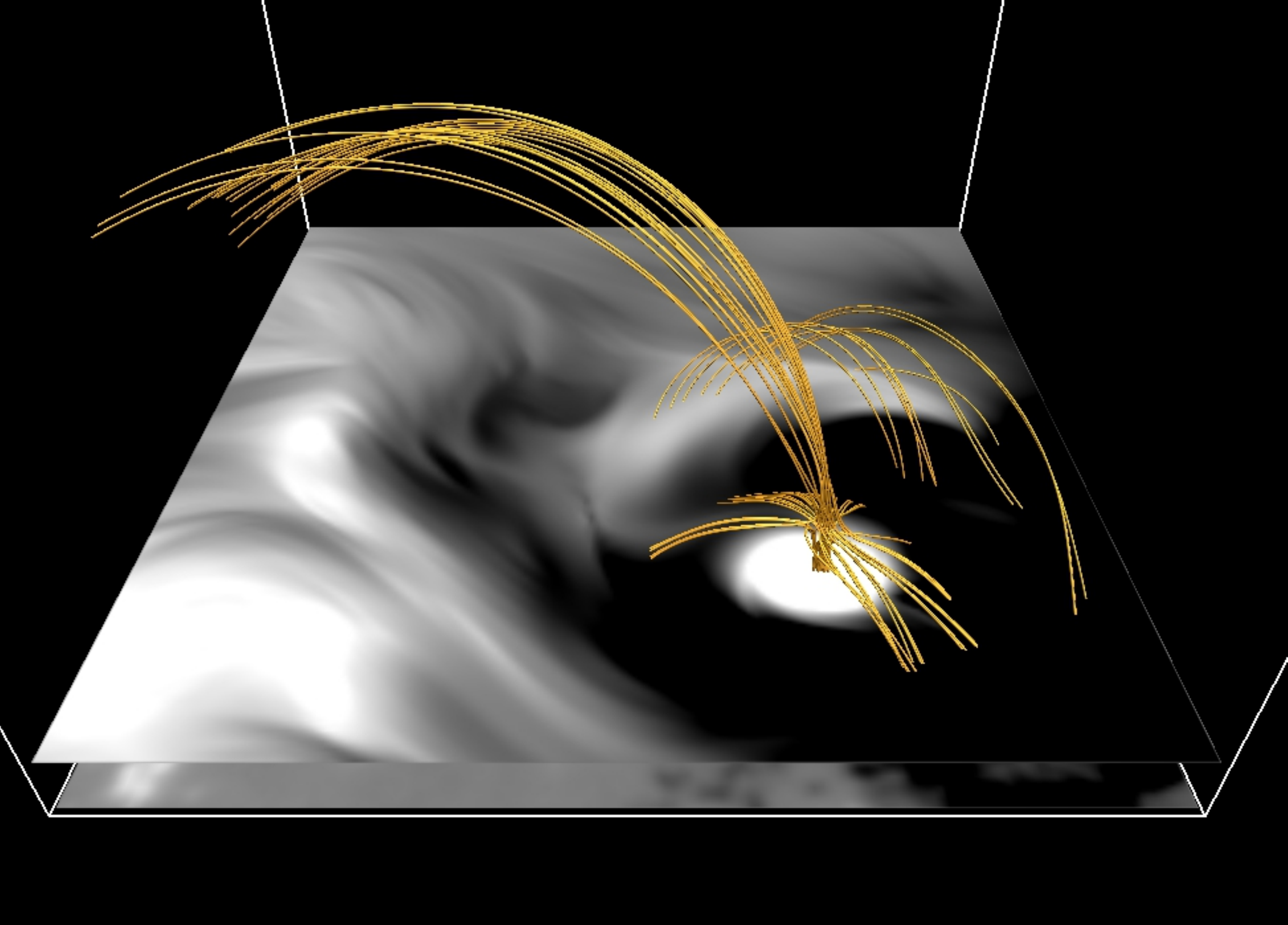}}\\
\subfigure[$B_z$ at $z=8$ Mm]{\includegraphics[width=0.4\textwidth]{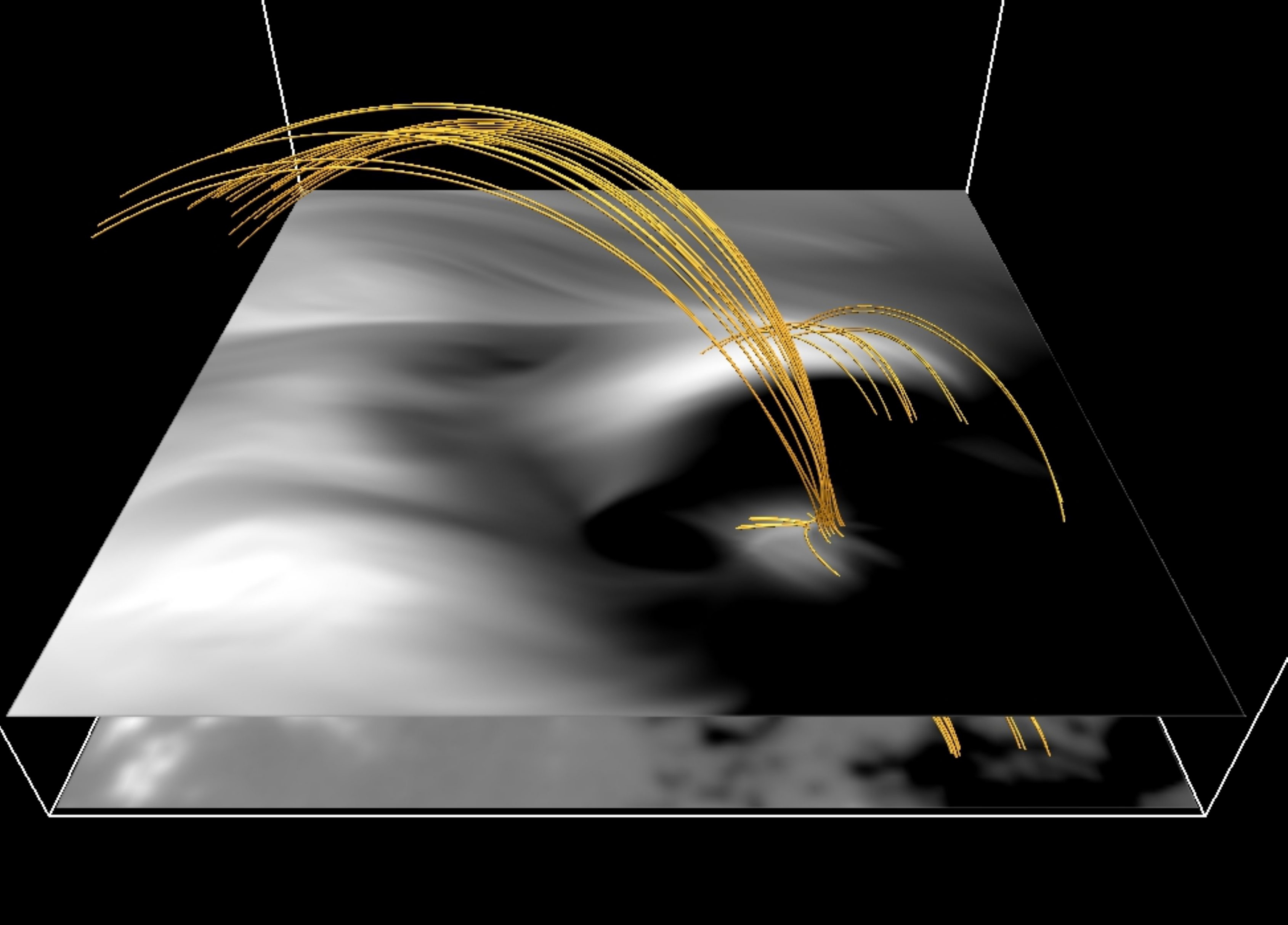}}
\caption{The magnetic configuration in the neighborhood of the parasitic pore in the data-driven model (with $U_0 = 1.1$ km s$^{-1}$) at time 11:02 UT.  Magnetic field lines traced from seed points in the vicinity of a coronal null point reveals a fan-spine topology, with the spine connecting the null point with the underlying parasitic polarity.}\label{fig8}
\end{figure}

Fig.~\ref{fig8} shows a 3D visualization of the magnetic structure from the same data-driven simulation ($U_0 = 1.1$ km s$^{-1}$) at 11:02 UT. The topological structure in this case is similar to those in scenarios examined by~\citet[][\citeyear{Pariat:2010Jet}]{Pariat:2009Jet}. In their numerical experiments, a parasitic polarity is embedded in an ambient field that is either vertical~\citep{Pariat:2009Jet} or inclined~\citep{Pariat:2010Jet}. In both cases, the model coronal field has a null point. Associated with the null point is a spine field line connecting the parasitic polarity with the null and a set of magnetic field lines forming a fan locus~\citep[i.e. a so-call fan-spine topology, see e.g.][]{Parnell:3DNulls}. As shown in Fig.~\ref{fig8}, the same type of magnetic skeleton is found in the data-driven model. In this case, the ambient field (with polarity opposite to that of the pore) is concentrated at the supergranular boundary, so the fan field lines emanating from the coronal null are connected to network flux. This type of topology is similar to the one inferred for the (un)twisting jet event studied by~\citet{Guo:RecurrentJets}~and~\citet{Schmieder:TwistingJet}. That event also occurred above a parasitic patch embedded in a supergranule at the edge of an \revision{AR}. 

\begin{figure}
\centering
\subfigure[\mcmc{Close-up view of magnetic flux rope at 12:55 UT}]{\includegraphics[width=0.4\textwidth]{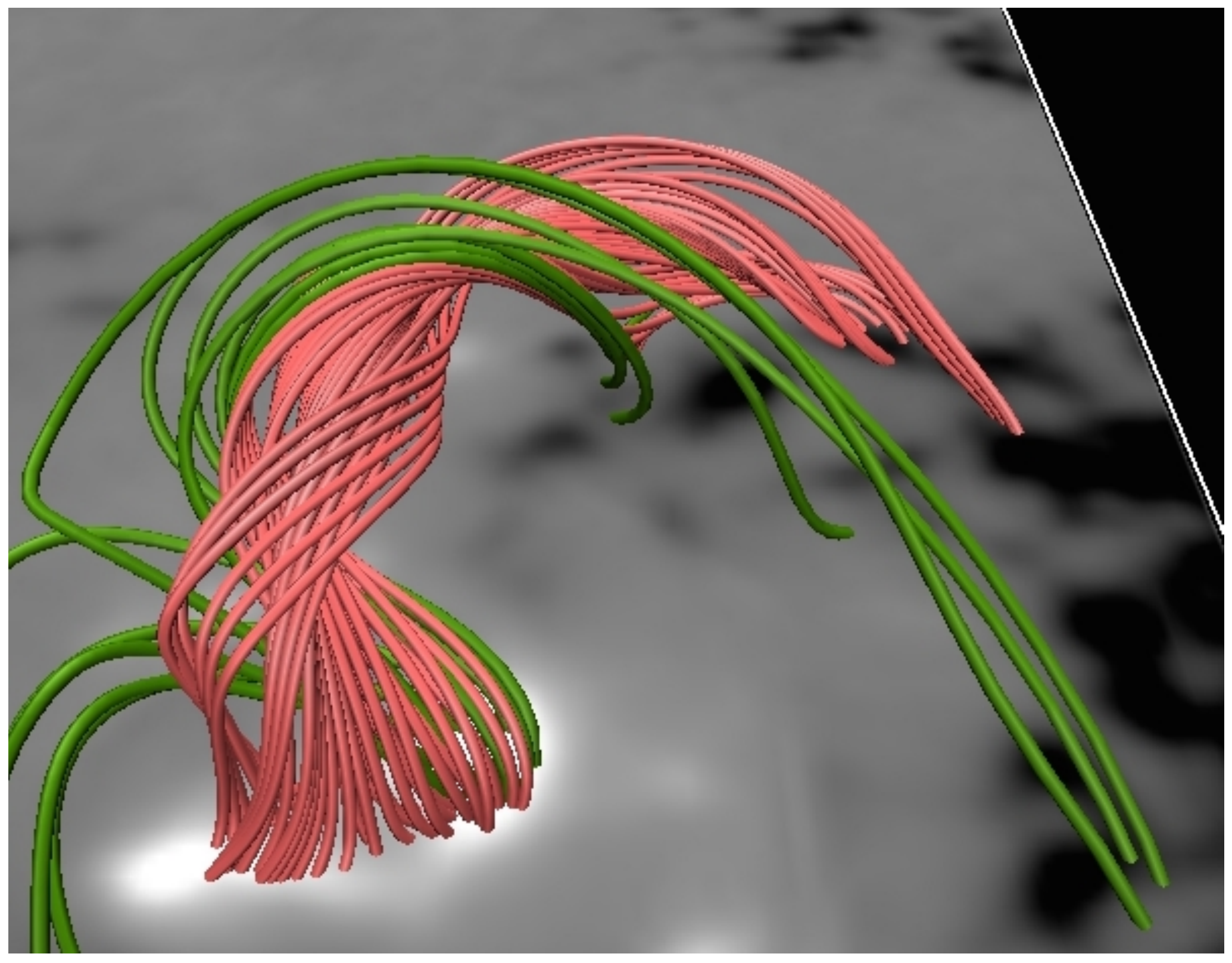}}\\
\subfigure[Magnetic field at 12:59 UT]{\includegraphics[width=0.4\textwidth]{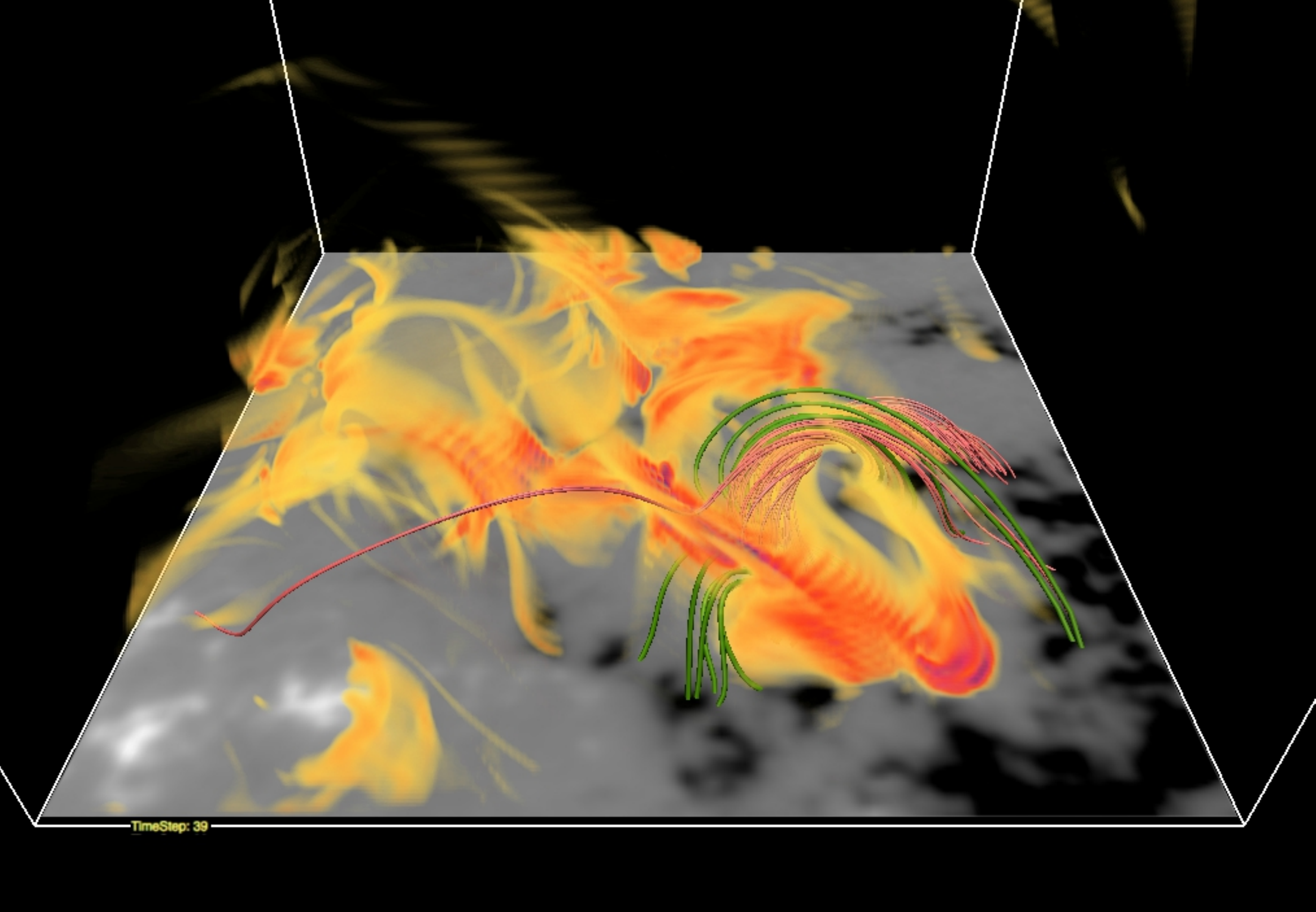}}\\
\subfigure[Magnetic field at 13:37 UT]{\includegraphics[width=0.4\textwidth]{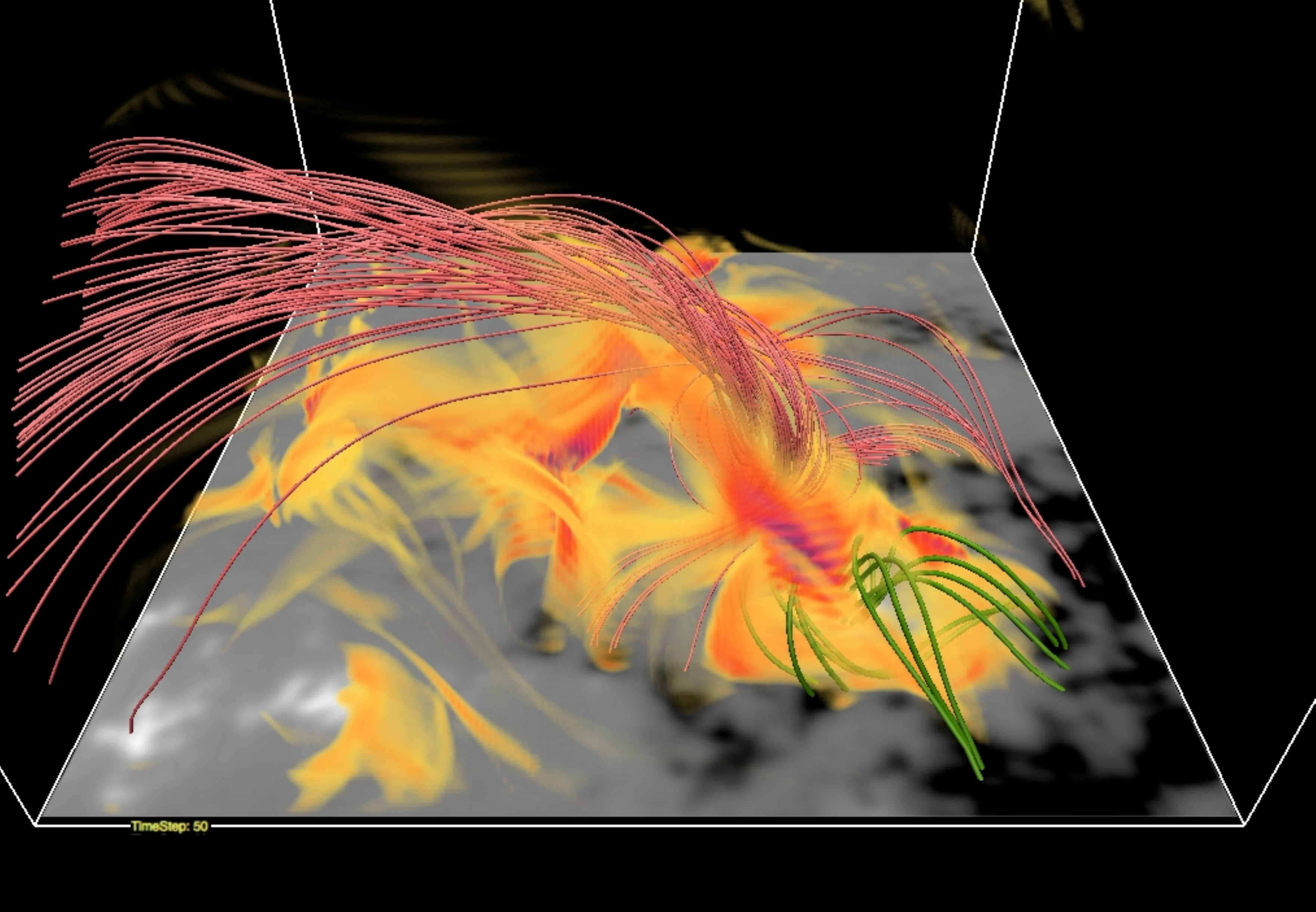}}
\caption{Jet-like magnetic evolution in the data-driven model (with $U_0=1.1$ km s$^{-1}$). Pink magnetic field lines are traced from a stationary grid of points. \mcmc{At 12:55 UT, they show a twisted flux rope structure}. At 13:37 UT, field lines traced from the same set of points reveal a set of inclined twisted field lines aligned with the background inclined field. Green field lines are traced from $z=0$ from the parasitic polarity. The semitransparent orange/red surfaces \mcmc{in panels (b) and (c)} indicate regions of strong current density.}\label{fig9}
\end{figure}

As demonstrated by the numerical MHD experiments by~\citet[][\citeyear{Pariat:2010Jet}]{Pariat:2009Jet}, persistent rotation of the parasitic polarity leads to a series of homologous helical jets. The simplistic MF model used in our data-driven simulations do not solve for the continuity, momentum nor energy equations so the model does not yield plasma ejections. However one can still examine the magnetic evolution in the data-driven model and find similarities between magnetofrictional evolution and MHD evolution. Fig.~\ref{fig9} shows two snapshots from a 3D visualization of the magnetic evolution in magnetofrictional model with $U_0=1.1$ km s$^{-1}$. \mcmc{At 12:55 UT} in the model, the pink field lines reveal a twisted flux rope connecting the parasitic pore with the northwestern edge of supergranule boundary. At 13:37 UT, the field has evolved so that the pink field lines (traced from the same positions) trace out a twisted bundle of inclined field.  The sign of magnetic twist in this flux bundle is consistent with the sign of rotation revealed by IRIS Doppler mean shift maps shown in Fig.~\ref{fig5}. That is, the sense of twist in this bundle results in a Lorentz force that drives rotational motion with $\omega_l < 0$ (see section~\ref{sec:tr_iris}).

In the~\citet{Pariat:2009Jet} model, the parasitic polarity was embedded in an ambient field that is purely vertical. For this setup, they found that the critical number of windings needed to be injected into the system to form a helical jet to be $N=1.4$~\citep[see also][]{Rachmeler:SymmetricCoronalJets}.  When the number of windings injected by surface rotation reached this value, the system underwent a helical kink instability, which broke the azimuthal symmetry of the system and generated a helical jet. The exact value of $N$ depends on the geometry of the system.  When the ambient open field is inclined, the azimuthal symmetry is no longer present and the critical threshold for injection of twist is lower, with a value of $N = 0.85\pm0.1$~\citep{Pariat:2010Jet}.

The average length of time required for twisting the field before jet-like reconfigurations occur in the data-driven models is consistent with the findings of~\citet{Pariat:2010Jet}. To estimate the windings injected per unit time in the data-driven simulations, consider an axisymmetric twisted flux tube with longitudinal and transverse components given by
\begin{eqnarray}
B_l(r)& =& B_0 e^{-r^2/R^2},\\
B_t(r)& = & \frac{\lambda r}{R} B_0 e^{-r^2/R^2}.
\end{eqnarray}
\noindent Here $r$ is the radial distance from the axis of symmetry, $\lambda$ is the dimensionless twist parameter and $R$ is the characteristic radius of the flux tube. This magnetic configuration consists of helical field lines that form concentric flux surfaces about the tube axis. The field lines have magnetic pitch such that the number of turns about the axis ($N$) over an axial distance $d$ is independent of radial distance $r$ and is given by:
\begin{equation}
N = d \frac{\lambda}{2 \pi R}.
\end{equation}
\noindent Consider the scenario in which magnetic twist is injected into the corona by a vertically aligned twisted flux tube rising through the photospheric layer ($z=0$) with speed $U_0$. Over a time period $\Delta t$, the number of turns injected is $N = U_0 \Delta t \lambda (2\pi R)^{-1}$. By inspecting the HMI vector magnetograms (fig.~\ref{fig:vmags}), we find that the horizontal components of the current-carrying field in the vicinity of the parasitic pore to be comparable to the vertical field strength inside the pore (approaching $1$ kG). So for rough estimates we can take $\lambda \sim 1$. Taking the size of the pore as $R\sim 1500$ km, we find that the time taken to reach the critical threshold of $N=0.85$ to be $\Delta t = 2.2 U_0^{-1}$ hr, where $U_0$ is in units of km s$^{-1}$. Within the 4-hour period in the simulations (10:00 UT and 14:00 UT), the number of jet-like reconfigurations in our data-driven simulations with $U_0 = 0$, $1.1$ and $2.2$ km s$^{-1}$ are 0, 2 and 4, respectively. So the amount of twisting required to drive jet-like episodes in the data-driven model is consistent with what is required in the MHD model of~\citet{Pariat:2010Jet}.

There are limitations to the magnetofrictional models. Although they give us physical guidance about how magnetic energy accumulates and how the magnetic field relaxes by unwinding, the amplitude of velocities in the models is dependent on the free magnetofrictional parameter $\nu_0$. For smaller values of $\nu_0$, the relaxation velocities have higher amplitude. The magnetofrictional models were not fine-tuned to reproduce the Doppler speeds in the observations (generally in our model the speeds are lower by a factor of a few). Furthermore, the relaxation velocity ($\vec{v} \propto \vec{j}\times\vec{B}$) is always perpendicular to $\vec{B}$, so the model cannot give predictions of the outflow speeds of jet material (from the IRIS slit-jaw image sequences, the outflow speeds of the jets reach beyond $100$ km s$^{-1}$). Another limitation is the lack of a treatment of thermodynamics quantities such as temperature and density. To overcome these limitations would require data-driven, fully-compressible MHD simulations to be performed in future studies. \mcmc{Another valuable exercise would be to compare the amount of twist injection needed in MHD simulations of flux emergence~\revision{\citep[those yielding helical jets,~e.g.][]{ArchontisHood:BlowoutJets,Moreno-Insertis:2013Jet,Fang:RotatingSolarJets,Lee:HelicalBlowoutJets}} with the results of~\citet{Pariat:2010Jet}.}

\section{Discussion}
\label{sec:discussion}
Over a four-hour period on July 21st 2013, recurrent jets emanating from NOAA AR 11793 were simultaneously observed by IRIS, SDO and Hinode. Doppler shift maps in the IRIS \ion{Si}{IV} 1394 \AA~transition region line shows all four jets exhibiting helical motion of the same sign. The IRIS Doppler shift maps share considerable resemblance to synthetic Doppler maps in~\citet{Fang:RotatingSolarJets}, who carried out MHD simulations of jets resulting from the interaction of a twisting flux tube emerging from the solar convection zone into a coronal layer with ambient inclined field.

Photospheric vector magnetograms from Hinode/SOT and SDO/HMI show that the source region of the homologous jets consists of predominantly negative polarity field concentrated at the boundary of a supergranule. Embedded inside the supergranule is a parasitic pore with positive magnetic flux. This type of photospheric flux distribution gives a coronal field with a spine-fan topology, which is common in 3D MHD models of coronal jets~\citep[e.g.][]{Moreno-Insertis:2008Jet}. 

Photospheric vector magnetograms from Hinode/SP and SDO/HMI show a persistent current-carrying magnetic configuration in the vicinity of the parasitic pore. Furthermore, the temporal sequence of vector magnetograms from SDO/HMI shows evidence for the emergence of magnetic flux in this current-carrying region (see section~\ref{sec:hmi_vmag}). 

To investigate the driving mechanism for the homologous helical jets, we performed a number of data-driven numerical simulations. All of the numerical simulations are driven by a bottom boundary condition that matches the evolution of the photospheric-$B_z$ as observed by HMI. \revision{The temporal sequence of $B_z$ indicates flux emergence is in progress during the time period when helical jets are observed.} The occurence of helical jet-like evolution in some simulation runs and not others implies (for this particular case) \mcmc{that the increase in unsigned magnetic flux ($|B_z|$) associated with emerging flux} is not a sufficient condition for helical jet formation.~\mcmc{What vector magnetograms (Fig.~\ref{fig:vmags}) reveal is that the emerging flux is current-carrying (i.e. has magnetic twist).  In the numerical experiment for which the driving electric fields at the bottom boundary are completely decoupled from the photospheric $j_z$ distribution, we find no helical, jet-like reconfigurations in the magnetic field model. In cases where twisting is imposed (i.e. the driving electric fields are coupled to $j_z$), the number of jet-like episodes within a 4-hour period increases linearly with the injection parameter $U_0$ (see Eq.~\ref{eqn:adhoc}). This suggests that the injection of twist via the emergence of current-carrying magnetic field is important for the creation of recurrent helical jets studied here. The amount of twist injection required between successive jet-like episodes is consistent with the findings of~\citet{Pariat:2010Jet}, who carried out fully-compressible MHD simulations to model the formation of homologous helical jets. In their numerical experiments, twist injection is due purely to rotational motion in the photosphere. However, MHD simulations of flux emergence from the convection zone into the atmosphere~\revision{\citep[e.g.][]{ArchontisHood:BlowoutJets,Moreno-Insertis:2013Jet,Fang:RotatingSolarJets,Lee:HelicalBlowoutJets}} also show signs of twist injection (by both bodily emergence of current-carrying field and Lorentz-force driven rotational motions) followed by emission of helical jets. It is likely both contributions are present for the observed helical jets studied here (which are found above a region of emerging flux).} 

This work is an example of how complementary observations from multiple observatories can be used in tandem with data-driven modeling to investigate the dynamics of the solar atmosphere. Co-spatial and simultaneous observations from IRIS, SDO and Hinode provide evidence for the helical nature of the recurrent jets and reveal the magnetic environment of their source region. The use of HMI vector magnetograms to perform data-driven simulations allowed us to investigate how processes such as flux emergence drive coronal evolution. The data-driven simulations were carried out with a magnetofrictional model, which is able to capture how a magnetic configuration relaxes in response to the Lorentz force. However the model lacks substantial physics and is not suitable for answering questions related to how the stored magnetic energy is used to heat previously cool plasma to transition region and coronal temperatures, and how the plasma ejected along the jet is accelerated. For example, an important question regarding jets is whether the ascending material is directly accelerated by the Lorentz force in reconnected field lines, or whether it is due to chromospheric evaporation, slow mode waves, or upward propagating shocks~\citep[e.g., see][]{Takasao:2013Jet}. Some of these outstanding issues will likely be addressed in the near future by more detailed analyses of IRIS spectra of chromospheric and transition region lines. Going forward, an improvement over the present study would involve data-driven, fully-compressible MHD simulations~\citep[see][for an example of data-driven MHD modeling applied to quasi-steady \revision{AR} coronal loops]{Bourdin:DataDrivenCorona}. Measurements of the vertical gradients of the magnetic field from vector magnetograms at two heights (photosphere and chromosphere) will help retrieve the driving electric field, but advances in deriving consistent boundary conditions in terms of the appropriate mass, momentum, and energy fluxes are also needed. The ability to do so will allow us to strengthen the constraints imposed by observations (e.g. in terms of temperature and density diagnostics) on theory and to better interpret observations based on realistic physical models.

\acknowledgements

Data are courtesy of the science teams of IRIS, SDO and Hinode. IRIS is a NASA small explorer mission developed and operated by LMSAL with mission operations executed at NASA Ames Research center and major contributions to downlink communications funded by the Norwegian Space Center (NSC, Norway) through an ESA PRODEX contract. This work is supported by NASA under contract NNG09FA40C (IRIS), the European Research Council grant agreement No. 291058, and contract 8100002705 from LMSAL to SAO. 

Additionally, MCMC acknowledges support from NASA's SDO/AIA (NNG04EA00C) and Hinode/SOT (NNM07AA01C) contracts and grants (NNX14AI14G and NNX13AJ96G) to LMSAL. AIA is an instrument onboard the Solar Dynamics Observatory, a mission for NASA's Living With a Star program. Hinode is a Japanese mission developed and launched by ISAS/JAXA, collaborating with NAOJ as domestic partner, and NASA and STFC (UK) as international partners. Science operation of Hinode is conducted by the Hinode science team organized at ISAS/JAXA. Postlaunch operation support is provided by JAXA and NAOJ (Japan), STFC (UK), NASA, ESA, and NSC (Norway).

The numerical modeling work is made possible by NASA's High-End Computing Program. The simulations presented in this paper were carried out on the Pleiades cluster at the Ames Research Center.

\bibliographystyle{apj}
\bibliography{refs,apj-jour}
\end{document}